\documentclass[12pt]{article} 
\usepackage[paper=letterpaper,margin=.8in]{geometry}

\usepackage{amsmath}
\usepackage{amsthm}
\usepackage{url}
\theoremstyle{plain}
\usepackage{amssymb}
\usepackage{tensor}
\usepackage{graphicx}
\usepackage{caption}
\usepackage{color}
\usepackage{subfigure}
\usepackage{tikz}
\usepackage[numbers,sort&compress]{natbib}
\usepackage[utf8]{inputenc}
\usepackage{braket}
\usepackage{upgreek}
\usepackage[normalem]{ulem}
\usepackage[colorlinks = true, linkcolor = blue, citecolor = red, urlcolor  = blue, anchorcolor = blue]{hyperref}
\newtheorem*{theorem*}{Theorem}
\newtheorem*{definition*}{Definition}

\newcommand{\calk}{\mathcal{K}}
\newcommand{\calo}{\mathcal{O}}
\newcommand{\bdyO}{O}
\newcommand{\calh}{\mathcal{H}}
\newcommand{\calt}{\mathcal{T}}

\newcommand{\cali}{\mathcal{I}}
\newcommand{\calw}{\mathcal{W}}
\newcommand{\cals}{\mathcal{S}}
\newcommand{\cala}{\mathcal{A}}
\newcommand{\calb}{\mathcal{B}}
\newcommand{\dt}{\tilde{\Delta}}

\numberwithin{equation}{section}

\begin{document}

	\thispagestyle{empty}

	\vspace*{2.5cm}
\begin{center}

{\bf {\LARGE von Neumann algebras in JT gravity}}\\

\begin{center}

\vspace{1cm}

{\bf David K. Kolchmeyer}\\
 \bigskip \rm

\bigskip 

Center for Theoretical Physics,\\Massachusetts Institute of Technology, Cambridge, MA 02139, USA

\rm
  \end{center}

\vspace{2.5cm}
{\bf Abstract}
\end{center}
\begin{quotation}
\noindent

We quantize JT gravity with matter on the spatial interval with two asymptotically AdS boundaries. We consider the von Neumann algebra generated by the right Hamiltonian and the gravitationally dressed matter operators on the right boundary. We prove that the commutant of this algebra is the analogously defined left boundary algebra and that both algebras are type II$_\infty$ factors. These algebras provide a precise notion of the entanglement wedge away from the semiclassical limit. We comment on how the factorization problem differs between pure JT gravity and JT gravity with matter.
\end{quotation}

	\pagebreak

	{
		\hypersetup{linkcolor=black}
		\tableofcontents
	}


	\pagebreak

	\section{Introduction}
	 
	The application of the theory of operator algebras to quantum gravity has led to various interesting results on the relationship between spacetime and quantum mechanics. The study of operator algebras was pioneered by von Neumann \cite{vN}, who was interested in developing the mathematics underlying quantum mechanics. A von Neumann (vN) algebra is a unital $*$-algebra of bounded operators that is closed in the weak topology. Physically, a vN algebra may be thought of as the algebra of observables that describes the possible measurements that may be performed on a quantum system. A fundamental result is the classification of vN algebra factors, which are the building blocks of more general vN algebras. A factor can be of types I, II, or III. A type I factor is equivalent to the algebra of all bounded operators acting on a Hilbert space, and is the only factor that arises in finite-dimensional systems. In infinite-dimensional systems, such as those that arise in statistical mechanics and quantum field theory, type II and III factors appear.\footnote{See \cite{Sorce:2023fdx} for a recent review of the type classification.}
	
	A basic kinematical question is how any given algebra of observables fits into this classification scheme. Understanding the types of vN algebras that appear can reveal many interesting properties of the system. For example, Leutheusser and Liu \cite{Leutheusser:2021frk,Leutheusser:2021qhd, Leutheusser:2022bgi} demonstrated the existence of emergent type III$_1$ factors in the large $N$ limit of a holographic CFT and used the theory of half-sided modular translations, which only exists for type III$_1$ factors, to make precise the notion of emergent bulk time. Furthermore, in a series of papers \cite{Witten:2021unn, Chandrasekaran:2022cip, Chandrasekaran:2022eqq}, Witten and collaborators precisely characterized the algebra of observables in the exterior of a two-sided AdS black hole and in the dS static patch in semiclassical gravity. They found that these algebras are type II factors, which immediately implies that the subsystem entanglement entropy is completely fixed up to a state-independent additive constant. Recently, Penington and Witten \cite{Penington:2023dql} reached a similar conclusion in the context of Jackiw-Teitelboim (JT) gravity \cite{Jackiw:1984je,Teitelboim:1983ux} minimally coupled to an arbitrary matter QFT. We expect that techniques from operator algebras will continue to improve our understanding of how gravitational physics emerges from holographic quantum systems.
	
	In this paper, we continue the work of \cite{Penington:2023dql} on the algebraic structure of JT gravity with matter. The dynamics of JT gravity is governed by the spontaneous breaking of the one-dimensional reparameterization group to $SL(2,\mathbb{R})$, which also occurs in other physically interesting systems like the SYK model or the near-horizon region of a near-extremal black hole \cite{Maldacena:2016upp, Maldacena:2016hyu}. JT gravity with matter is exactly solvable, and interesting vN algebras appear at both weak and strong coupling. In particular, in the semiclassical (or weak-coupling) limit, one may define emergent type III$_1$ factors in the canonical ensemble (as in \cite{Leutheusser:2021frk,Leutheusser:2021qhd}) or emergent type II$_\infty$ factors in the microcanonical ensemble (as in \cite{Chandrasekaran:2022cip}). As in \cite{Penington:2023dql}, our focus will be on the strong-coupling regime, where we study the vN algebras associated to the left and right AdS boundaries of a two-sided black hole. We are ultimately interested in how semiclassical algebras arise from nonperturbative bulk dynamics in general holographic systems, and JT gravity serves as a toy model where the algebras can be explicitly described both before and after the semiclassical limit is taken. We define the right boundary algebra $\cala_R$ to be generated by the right Hamiltonian and all of the matter operators on the right boundary. The left algebra $\cala_L$ is defined similarly. In the context of higher-dimensional AdS/CFT, the matter operators should be thought of as the single-trace operators dual to bulk matter fields and their (normal-ordered) products. In the semiclassical limit, after $N$ is taken to $\infty$, these operators generate the bulk causal wedge via HKLL reconstruction \cite{Hamilton:2005ju}. In contrast, in the strongly-coupled regime, we prove that $\cala_L$ and $\cala_R$ are commutants of each other, meaning that they behave more like entanglement wedges than causal wedges. We also prove that $\cala_L$ and $\cala_R$ are factors. These properties strongly suggest that $\cala_L$ and $\cala_R$ generalize the entanglement wedge away from the semiclassical limit. It follows that the algebra of boundary operators that have a good semiclassical limit is much larger than the algebra of operators that generate the causal wedge. This reinforces the claim of \cite{Leutheusser:2022bgi} that modular-flowed single-trace operators also have a good large $N$ limit and are necessary for generating the entanglement wedge. Furthermore, note that the union of $\cala_L$ and $\cala_R$ generates the entire type I algebra of bounded operators on the Hilbert space \cite{Penington:2023dql}. This includes interesting two-sided observables such as (bounded functions of) the wormhole length operator. The main difference between the boundary algebras $\cala_L$ and $\cala_R$ and a pair of holographic CFTs at finite $N$ is that the CFT algebras are type I factors, whereas the JT gravity boundary algebras are type II$_\infty$ factors. Despite this factorization problem \cite{Harlow:2018tqv}, we believe that strongly-coupled JT gravity with matter can be useful for demonstrating the emergence of semiclassical physics.
	
	\begin{figure}
	\centering
	\includegraphics[width=0.7\linewidth]{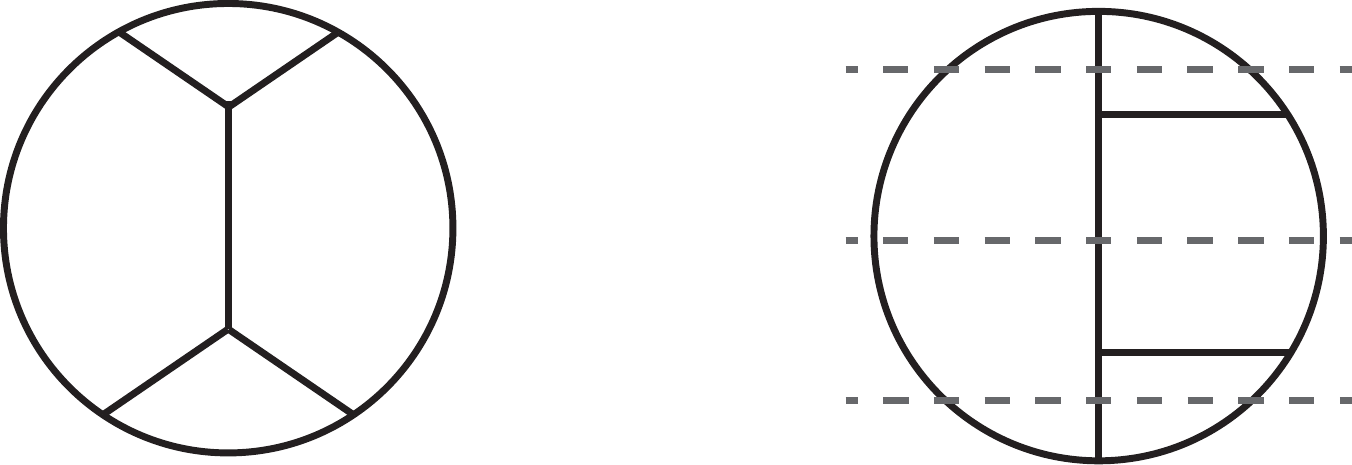}
	\caption{The left diagram depicts a four-point conformal block for a matter theory quantized on the hyperbolic disk. The matter operators are inserted on the boundary of the disk. By applying a simple set of rules to this diagram (and summing over the exchanged states/operators), we may obtain an expression for the disk four-point function in JT gravity coupled to matter. On the right, we separate the diagram into sub-diagrams to represent it as an expectation value of operators that act on a Hilbert space. The bottom sub-diagram represents a ket that corresponds to the Hartle-Hawking state with a matter operator insertion, and each of the two middle sub-diagrams represents an operator inserted on the right boundary. The top sub-diagram is the bra.} 
		\label{fig:fourpointintro}
\end{figure}
	
	To analyze the boundary algebras, we represent their generators with diagrams. This is similar in spirit to how diagrams are used to represent, for example, the generators of the Temperley–Lieb algebra. In particular, we will show that a Euclidean disk $n$-point correlation function of matter operators may be computed by applying a set of rules to a tree graph with trivalent vertices drawn inside a circle (see Figure \ref{fig:fourpointintro}). We may separate such a diagram into sub-diagrams that separately represent a bra, a ket, and the operators that are inserted in between. The tree graph represents a conformal block, and the diagrammatic rules depend on the three-point structure constants of the matter QFT. Associativity of the OPE guarantees that all of the different possible tree graphs produce the same result for the correlator (after summing over the exchanged primaries). Our diagrammatic techniques extend the work of \cite{Suh:2020lco, Mertens:2017mtv, Jafferis:2022wez}.
	
Our main new technical result relative to \cite{Penington:2023dql} is a proof that the boundary algebras are factors, as well as an alternate proof that the two algebras are commutants of each other. Whereas \cite{Penington:2023dql} used Euclidean methods to prove the latter statement, our proof is Lorentzian in nature, as it only uses the expressions for the matrix elements of the generators in the energy basis. We also provide a new example of a positive densely-defined trace.

	In section \ref{sec:review}, we review JT gravity and the boundary particle formalism. In section \ref{sec:matterondisk}, we summarize aspects of QFT on AdS$_2$ for the purpose of setting conventions. In section \ref{sec:pathintegralquant}, we quantize JT gravity with matter by rewriting Euclidean disk correlators as inner products. In section \ref{sec:algebras}, we define $\cala_R$ and $\cala_L$ and prove their interesting properties. In section \ref{sec:tracesection}, we reiterate the existence of a unique trace (up to an overall rescaling) on these algebras and introduce a weaker notion of the trace that still preserves some physically important properties but is not unique. In section \ref{sec:facproblem}, we comment on the implications of our results for attempts to factorize the Hilbert space using edge modes, emphasizing the differences between pure JT gravity and JT gravity with matter. Further directions are summarized in section \ref{sec:discussion}.
	
	\section{Review of JT gravity}
	
	\label{sec:review}
	
	In this section, we review the aspects of JT gravity that we will need later. Readers who want more details and references are encouraged to read \cite{Mertens:2022irh}. The Euclidean action of JT gravity minimally coupled to matter is
	\begin{align}
	I[g,\Phi,\varphi] &= - S_0 \chi  + I_{JT}[g,\Phi] + I_{m}[g,\varphi] \ ,  
	\\
	I_{JT}[g,\Phi] &= -  \int_{\mathcal{M}} \sqrt{g} \Phi (R + 2) - 2\int_{\partial \mathcal{M}} \sqrt{h} \Phi (K-1) \ ,
\end{align}
where we have set $16 \pi G_N = 1$, and $I_m[g,\varphi]$ is the action of a matter QFT. The Euler characteristic $\chi$ and the zero-temperature entropy $S_0$ will not play an important role in this paper, because we mainly focus on disk topologies. On the AdS boundary, we impose that the component of the metric along the boundary obeys
\begin{equation}
\left.g_{\tau \tau}\right|_{\partial \mathcal{M}} = \frac{1}{\epsilon^2},
\end{equation}
while the boundary value of the dilaton is fixed to be
\begin{equation}
\left. \Phi \right|_{\partial \mathcal{M}} = \frac{\phi_b}{\epsilon}.
\end{equation}
We always work in the limit $\epsilon \rightarrow 0$. The parameter $\phi_b$ has dimensions of length and sets the overall scale of the correlation functions. The role of $G_N$ is played by $\frac{1}{\phi_b}$, which means that the semiclassical (or weak-coupling) limit, $\phi_b \rightarrow \infty$, is a high-energy limit that is analogous to the thermodynamic limit in statistical systems \cite{Jafferis:2022uhu}, or the large $N$ limit in holographic CFTs.
	
\begin{figure}
	\centering
	\includegraphics[width=0.2\linewidth]{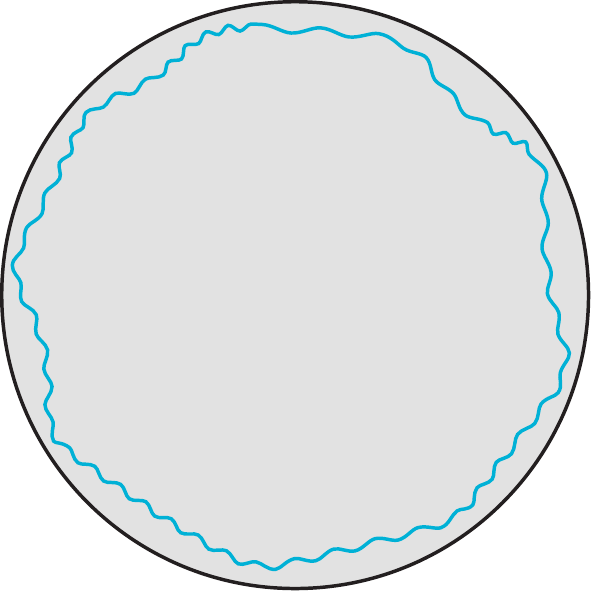}
	\caption{The gray disk represents the hyperbolic disk, and the blue curve is the trajectory of the boundary particle, or the boundary of the physical manifold. The disk Euclidean path integral of JT gravity integrates over all geometries that can be realized as the interior of the blue curve. The condition $\phi(\tau + \beta) = \phi + 2 \pi$ ensures that the boundary curve wraps once around the disk.}
	\label{fig:diskwiggly}
\end{figure}

The remainder of this section is dedicated to a review of the boundary particle formalism \cite{Yang:2018gdb,Kitaev:2018wpr,Suh:2020lco} for computing the Euclidean path integral. See also the review in \cite{Lin:2022zxd}. Our main focus is on disk correlators. The boundary particle is defined using the following quantum-mechanical path integral,\footnote{Throughout this paper the $\prime$ superscript refers to a $\tau$ derivative. See appendix B of \cite{Jafferis:2019wkd} for more comments on this path integral.}
	\begin{equation}
		\label{eq:Euclideanschwarzianpathintegral}
		\begin{split}
			&K_\beta(\phi_2,\psi_2,\phi_1,\psi_1) :=
			\\
			&\int \mathcal{D}\phi \mathcal{D}\psi \mathcal{D}\pi_\phi \mathcal{D}\pi_\psi  \exp\left(\int_0^\beta d\tau \left[ i \pi_\psi \psi^\prime +  i \pi_\phi \phi^\prime - \frac{1}{2 \phi_b} \left[ \frac{\pi_\psi^2}{2} + i \pi_\phi e^\psi - \frac{1}{2} e^{2 \psi}\right]   \right]\right),
		\end{split}
	\end{equation}
which may be viewed as a worldline theory of a particle whose target space is the asymptotic region of the Euclidean hyperbolic disk. The $\phi$ field refers to an angular coordinate on the Euclidean disk, while $\psi$ is a radial coordinate. Although $\phi$ is interpreted as an angular coordinate, we choose to let $\phi$ be a noncompact bosonic field in this path integral. The fields $\pi_\psi$ and $\pi_\phi$ are canonically conjugate to $\psi$ and $\phi$. The canonical Hamiltonian $H$ may be simply read from \eqref{eq:Euclideanschwarzianpathintegral},
\begin{equation}
	2 \phi_b H = \frac{\pi_\psi^2}{2} + i \pi_\phi e^\psi - \frac{1}{2} e^{2 \psi},
\end{equation}
and \eqref{eq:Euclideanschwarzianpathintegral} may be canonically rewritten as follows:
\begin{equation}
	K_\beta(\phi_2,\psi_2,\phi_1,\psi_1) = \braket{\phi_2 \, \psi_2|e^{-\beta H}|\phi_1 \, \psi_1},
\end{equation}
where
\begin{equation}
	\braket{\phi_2 \, \psi_2|\phi_1 \, \psi_1} = \delta(\phi_2 - \phi_1) \delta(\psi_2 - \psi_1).
\end{equation}
The trajectory of this particle should be interpreted as the location of the boundary of the target-space manifold. Put differently, because the off-shell path integral in JT gravity is restricted to manifolds with hyperbolic geometry, different off-shell configurations can be viewed as cutouts of a given two-dimensional hyperbolic manifold. The particle's trajectory is the location of the cut. See Figure \ref{fig:diskwiggly} for an illustration. 

Furthermore, the field $\pi_\phi$ acts as a Lagrange multiplier that enforces the constraint
\begin{equation}
	\phi^\prime = \frac{e^\psi}{2 \phi_b} > 0,
\end{equation}
which implies that the particle's trajectory is non-self-intersecting, since $\phi$ increases monotonically with the particle's time $\tau$.

The time $\tau$ should be viewed as the physical time of the boundary (for instance, it could represent Euclidean time in the SYK model). To better understand the asymptotic radial coordinate $\psi$, we may write the metric on the disk as follows,
\begin{equation}
	ds^2 = d\rho^2 + d\phi^2 \sinh^2 \rho,
\end{equation}
so that the trajectory of the particle is given by $\phi(\tau)$ and $\rho(\tau)$. The JT gravity boundary condition on the metric component $g_{\tau \tau} = \frac{1}{\epsilon^2}$ implies that
\begin{equation}
	\epsilon^2(\rho^\prime(\tau))^2 + (\phi^\prime(\tau))^2 \epsilon^2\sinh^2 \rho (\tau) = 1,
\end{equation}
and as $\epsilon \rightarrow 0$, $\rho(\tau)$ becomes large while $\rho^\prime(\tau)$ remains order 1. For small $\epsilon$, we then have that
\begin{equation}
\frac{e^{\psi(\tau)}}{2 \phi_b} =	\phi^\prime(\tau) \approx \frac{2}{\epsilon}  e^{-\rho(\tau)},
\label{eq:2.7}
\end{equation} 
which clarifies the role of $\psi$ as a radial coordinate and also shows that in our conventions, larger values of $\psi$ correspond to smaller radial distances.

If we integrate out all of the fields except for $\phi$ in \eqref{eq:Euclideanschwarzianpathintegral}, the result is
\begin{equation}
	\label{eq:swpathintegral}
	\int d\mu[\phi] \exp\left(-2 \phi_b \int_0^\beta d\tau \, \left(  \frac{1}{2} \left(\frac{\phi^{\prime \prime}}{\phi^\prime}\right)^2  - \frac{1}{2} ( \phi^\prime)^2  \right)\right),
\end{equation} 
which is the same path integral that was studied in \cite{Stanford:2017thb}. In this description, it is clear that $\phi_b \rightarrow \infty$ is a weak-coupling limit. The measure $d\mu[\phi]$ is given by
\begin{equation}
	d\mu[\phi] = \prod_\tau \frac{d\phi(\tau)}{\phi^\prime(\tau)}, \label{eq:2.9}
\end{equation}
which is invariant under reparameterizations of $\phi$. It follows that the Euclidean path integral on the disk may be computed as follows:
\begin{equation}
	Z_{\text{disk}}(\beta) = \frac{1}{\text{Vol } \widetilde{SL}(2,\mathbb{R})}\int_{-\infty}^\infty d\psi \int_{-\infty}^\infty d\phi \, K_\beta(\phi + 2 \pi,\psi,\phi,\psi).
	\label{eq:formaldisk}
\end{equation}
To obtain a finite result, we have formally divided the path integral by the infinite volume of $\widetilde{SL}(2,\mathbb{R})$, which is the universal cover of $PSL(2,R)$, the isometry group of the hyperbolic disk. This expression correctly computes the disk because we have explicitly enforced that the boundary particle trajectories obey $\phi(\tau + \beta) = \phi(\tau) + 2 \pi$. A periodicity other than $2 \pi$ corresponds to a defect inside the disk. We now discuss the symmetries of the path integral so that we can gauge fix \eqref{eq:formaldisk} more carefully.

\subsection{Symmetries}

Our conventions for the generators of $\mathfrak{sl}(2,\mathbb{R})$ are as follows:
\begin{equation}
\label{eq:matrixrep}
	T_1 = \left(
	\begin{array}{cc}
		0 & -\frac{1}{2} \\
		\frac{1}{2} & 0 \\
	\end{array}
	\right), \quad T_2 = \left(
	\begin{array}{cc}
		\frac{1}{2} & 0 \\
		0 & -\frac{1}{2} \\
	\end{array}
	\right), \quad T_3 = \left(
	\begin{array}{cc}
		0 & \frac{1}{2} \\
		\frac{1}{2} & 0 \\
	\end{array}
	\right).
\end{equation}
We may parameterize a group element $g \in \widetilde{SL}(2,\mathbb{R})$ in a patch of the group manifold as follows:
\begin{equation}
	g = e^{ T_1 \phi } e^{- \psi T_2} e^{v(T_1 - T_3)},
\end{equation}
where $\phi,\psi,v$ are coordinates. Given $h \in \widetilde{SL}(2,\mathbb{R})$, we define the transformed coordinates $(\phi^h,\psi^h,v^h)$ using the following equation:
\begin{equation}
hg := e^{ T_1 \phi^h } e^{- \psi^h T_2} e^{v^h (T_1 - T_3)}.
\end{equation}
Upon composing two transformations, we find that
\begin{equation}
	((\phi^{g_1})^{g_2},(\psi^{g_1})^{g_2},(v^{g_1})^{g_2}) = ( \phi^{g_2 g_1},\psi^{g_2 g_1},v^{g_2 g_1}), \quad \quad g_1,g_2 \in \widetilde{SL}(2,\mathbb{R}).
\end{equation}
If we define
\begin{equation}
	\delta_a x = \left. \frac{d}{ds}x^{e^{s T_a}} \right|_{s = 0}, \quad a \in \{1,2,3\}, \quad x \in \{\phi,\psi,v\},
\end{equation}
then the infinitesimal transformations are given by
\begin{align}
	\label{eq:first}
	\delta_1 \phi &= 1 , \quad \delta_1 \psi = 0 , \quad \delta_1 v = 0 , \\
		\label{eq:second}
	\delta_2 \phi &= - \sin \phi , \quad \delta_2 \psi = - \cos \phi , \quad \delta_2 v = e^\psi \sin \phi , \\
		\label{eq:third}
	\delta_3 \phi &= \cos \phi , \quad \delta_3 \psi = - \sin \phi , \quad \delta_3 v = - e^\psi \cos \phi . 
\end{align}
Despite the notation, $v^h$ depends not only on $v$ and $h$ but also on $\psi$ and $\phi$. Likewise, $\psi^h$ depends on $\phi$, $\psi$, and $h$. We leave these additional dependencies implicit in the notation.

Using the angular and radial coordinate interpretations of $\phi$ and $\psi$ discussed above, the infinitesimal transformations on $\phi$ and $\psi$ are the ones generated by the isometries of the Euclidean hyperbolic disk. Their transformations do not depend on $v$, which is an auxiliary coordinate that is sometimes convenient to keep in mind. Next, we define
\begin{equation}
g_1 = e^{ T_1 \phi_1 } e^{- \psi_1 T_2} e^{v_1(T_1 - T_3)}, \quad g_2 = e^{ T_1 \phi_2 } e^{- \psi_2 T_2} e^{v_2(T_1 - T_3)},
\end{equation}
and it follows that any function of the matrix elements of $g_2^{-1} g_1$ (using the matrix representation \eqref{eq:matrixrep}) is invariant. After inspecting the matrix elements of $g_2^{-1} g_1$, we define two invariant quantities:
\begin{align}
e^{-\ell_{12}/2} &:= \frac{ e^{\frac{\psi_1 + \psi_2}{2}}}{2\sin \left(\frac{\phi_2 - \phi_1}{2}\right)},
\label{eq:length}
\\
m_{12} &:=  (e^{\psi_2} + e^{\psi_1}) \cot \frac{\phi_1 - \phi_2}{2} + v_2 - v_1.
\label{eq:other}
\end{align}
In these definitions, we assume that $\phi_2 - \phi_1 \in (0,2 \pi)$. Note that $\ell_{12}$ is the invariant renormalized length between $(\phi_1,\psi_1)$ and $(\phi_2,\psi_2)$ on the hyperbolic disk. The renormalized length is equal to the bare length minus $2 \log \frac{2 \phi_b}{\epsilon}$. Throughout this paper, a length that is renormalized by subtracting $2 \log \frac{2 \phi_b}{\epsilon}$ will be denoted by an $\ell$, perhaps with a subscript and/or superscript. Such lengths correspond to geodesics that start and end in the asymptotic infinity region of the hyperbolic disk. Also, a length that is renormalized by subtracting $\log \frac{2 \phi_b}{\epsilon}$ will be denoted by a $w$, perhaps with a subscript and/or superscript. These lengths correspond to geodesics that start in the asymptotic infinity region of the disk but terminate in the bulk.

Note that the integration measure
\begin{equation}
	d\phi \, d\psi \,  e^{-\psi}
\end{equation}
is invariant under the transformations on $\phi$ and $\psi$.

\subsection{Boundary particle propagator}

We are finally ready to present the explicit expression for the boundary particle propagator, \eqref{eq:Euclideanschwarzianpathintegral}. For $\phi_2 - \phi_1 \in (0,2 \pi)$, it is 
\begin{equation}
	\label{eq:Euclideanschwarzianpropagator}
	\begin{split}
		K_\beta(\phi_2,\psi_2,\phi_1,\psi_1) = \exp\left((e^{\psi_2} + e^{\psi_1})\cot\left(\frac{\phi_1 - \phi_2}{2}\right) \right) \\
		\times  \frac{1}{\pi^2} \frac{1}{\sin \frac{\phi_2 - \phi_1}{2}} \int_{0}^{\infty}ds s \sinh(2 \pi s) e^{-\frac{s^2}{4 \phi_b}\beta} K_{2is}\left(\frac{ 2 e^{\frac{\psi_2 + \psi_1}{2}}}{\sin \frac{\phi_2 - \phi_1}{2}}\right).
	\end{split}
\end{equation}
As explained in appendix B of \cite{Jafferis:2019wkd}, this formula was directly lifted from \cite{Yang:2018gdb}, up to changes of conventions. Because we prefer to work with a symmetry-invariant expression, we define
\begin{align}
	\label{eq:invarfirst}
	\calk_{\beta,12} &:= e^{\frac{\psi_2}{2} + v_2}K_\beta(\phi_2,\psi_2,\phi_1,\psi_1) e^{\frac{\psi_1}{2} - v_1}, \\
	\label{eq:invarsecond}
	&= \frac{2}{\pi^2} e^{m_{12}-\ell_{12}/2}  \int_{0}^{\infty}ds s \sinh(2 \pi s) e^{-\frac{s^2}{4 \phi_b}\beta} K_{2is}\left(4 e^{-\ell_{12}/2}\right).
\end{align}
The notation $\calk_{\beta,12}$ is convenient because it does not refer to a specific way of choosing coordinates on the disk. If we need to explicitly refer to the coordinates, we will write $\calk_\beta(\phi_2,\psi_2,v_2 ; \phi_1,\psi_1,v_1)$. The extra $v$ variable will always drop out of our final expressions. The gluing property of this propagator follows from \cite{Yang:2018gdb} and appendix B of \cite{Jafferis:2019wkd}:
\begin{equation}
\begin{split}
&\int_{-\infty}^\infty d\psi_2 \, e^{- \psi_2} \int_{\phi_1}^{\phi_3} d\phi_2 \, 	\calk_{\beta_2}(\phi_3,\psi_3,v_3 ; \phi_2,\psi_2,v_2) \calk_{\beta_1}(\phi_2,\psi_2,v_2 ; \phi_1,\psi_1,v_1) 
\\
&= \int_{-\infty}^\infty d\psi_2 \, e^{- \psi_2} \int_{\phi_1}^{\phi_3} d\phi_2 \, 	\, \calk_{\beta_2,23}\calk_{\beta_1,12}
= \calk_{\beta_1 + \beta_2}(\phi_3,\psi_3,v_3 ; \phi_1,\psi_1,v_1).
\end{split}
\end{equation}

The readers interested in the origin of this propagator are encouraged to read \cite{Yang:2018gdb,Kitaev:2018wpr,Lin:2022zxd}. Instead of repeating their derivations, we will instead demonstrate the correctness of \eqref{eq:Euclideanschwarzianpropagator} by reproducing the known results for the disk \cite{Stanford:2017thb}, the disk with a conical defect \cite{Mertens:2019tcm}, and the trumpet \cite{Saad:2019lba}.

\subsection{Disk with a conical defect}

The easiest case to consider is the disk with a conical defect, because the isometry group is simply $U(1)$. We simply repeat \eqref{eq:formaldisk} but with a different periodicity $2 \pi - \alpha$:
\begin{equation}
\frac{1}{\text{Vol } U(1)}\int_{-\infty}^\infty d\psi \int_{0}^{2 \pi - \alpha} d\phi \, K_\beta(\phi + 2 \pi - \alpha,\psi,\phi,\psi),
\label{eq:2.26}
\end{equation}
and note that $\text{Vol } U(1) = 2 \pi - \alpha$ cancels the contribution from the $\phi$ integration. After exchanging the $\psi$ integration above with the $s$ integration in \eqref{eq:Euclideanschwarzianpropagator} and using 6.621.3 in \cite{grad}, \eqref{eq:2.26} becomes
\begin{equation}
\frac{1}{2 \sin \frac{\alpha}{2}}	\frac{1}{\pi} \int_0^\infty ds \,  \cosh(2 \pi s (1 - \alpha)) e^{-\frac{s^2}{4 \phi_b} \beta}. 
\end{equation}
This agrees with the result in \cite{Mertens:2019tcm} up to a factor of $2 \sin \frac{\alpha}{2}$. The discrepancy can be accounted for by comparing the measure used in \cite{Mertens:2019tcm} with the measure in \eqref{eq:Euclideanschwarzianpathintegral}, which becomes \eqref{eq:2.9} after integrating out all fields but $\phi$. The measure used in \cite{Mertens:2019tcm} is a canonical measure that one can define on a coadjoint orbit of the Virasoro group. Using the methods of section 2.2 of \cite{Stanford:2017thb}, one can explicitly show that the measure used in \cite{Mertens:2019tcm} differs from \eqref{eq:2.9} by $2 \sin \frac{\alpha}{2}$. Hence, the boundary particle formalism is consistent with prior calculations of the disk with a conical defect.

\subsection{Trumpet}

\label{sec:trumpet}
\begin{figure}
	\centering
	\includegraphics[width=0.4\linewidth]{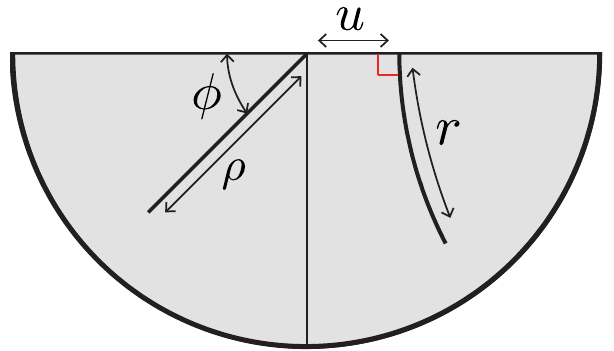}
	\caption{The $(\rho,\phi)$ coordinate system defined here is used throughout this paper. The $(r,u)$ coordinate system defined here is used in section \ref{sec:trumpet}. The thick black lines are geodesics with lengths $\rho$ and $r$.}
	\label{fig:coordconventions}
\end{figure}

We now use the boundary particle propagator to compute the trumpet. To do this, we must change coordinates to a different set of coordinates that is better suited for representing the trumpet as a subregion of the disk (with identifications). Consider the following hyperbolic metric on the half-disk:
\begin{equation}
	ds^2 = dr^2 + du^2 \cosh^2 r, \quad r \in (0,\infty), \quad u \in (-\infty,\infty).
\label{eq:2.28}
\end{equation}
If $r$ had the range $(-\infty,\infty)$, then these coordinates would fully cover the entire disk. Our coordinate conventions for the half-disk are shown in Figure \ref{fig:coordconventions}.
We define the coordinate $w$ by
\begin{equation}
	w := r - \log \frac{2 \phi_b}{\epsilon},
\end{equation}
so that $w$ and $u$ are coordinates of the asymptotic infinity region of the disk. The coordinate change between these coordinates and the $(\phi,\psi)$ coordinates is given as follows:
\begin{align}
	e^{-\psi} &= \frac{e^w}{2} \cosh u, \\
	- \cot \left( \frac{\pi + 2 \phi}{4}\right) &= \tanh \frac{u}{2},
\end{align}
and the invariant quantities $\ell_{12}$ and $m_{12}$ become
\begin{align}
	\label{eq:invar1t}
	 e^{-\ell_{12}/2} &= \frac{e^{\frac{-w_1 - w_2}{2}}}{\sinh \frac{u_2 - u_1}{2}}
	,
	\\
	\label{eq:invar2t}
	m_{12} &= - 2 (e^{-w_1} + e^{-w_2}) \coth \left( \frac{u_2 - u_1}{2}\right) + v_2 - v_1.
\end{align}
The invariant measure $d\phi \, d\psi \, e^{-\psi}$ becomes
\begin{equation}
	d\phi \, d\psi \, e^{-\psi} = \frac{du \, dw}{2} e^w.
\end{equation}
A trumpet geometry emerges from \eqref{eq:2.28} when we make the identification $u \sim u + b$, where $b$ is the length of the closed geodesic that the trumpet ends on. We define the trumpet propagator \begin{equation}\calk^t_{\beta}(u_2,w_2;u_1,w_1)
\end{equation} to be \eqref{eq:invarsecond}, with $m_{12}$ and $\ell_{12}$ given in \eqref{eq:invar1t} and \eqref{eq:invar2t} and $v_1 = v_2 = 0$. The trumpet is then given by
\begin{equation}
	\label{eq:trumpetcalc}
	\frac{1}{\text{Vol }U(1)}\int_{-\infty}^\infty \frac{dw \, e^{w}}{2} \int_0^b du \, \calk^t_{\beta}(u + b,w;u,w) = \int_{-\infty}^\infty \frac{dw \, e^{w}}{2}  \calk^t_{\beta}(b,w;0,w),
\end{equation}
where in this case $\text{Vol }U(1) = b$. Proceeding as in the previous subsection, we find that the trumpet becomes
\begin{equation}
	\frac{1}{2 \sinh \frac{b}{2}} \frac{1}{\pi} \int_0^\infty ds \, \cos(b s) e^{-\frac{s^2}{4 \phi_b} \beta}.
\label{eq:trumpet}
\end{equation}
In complete analogy to the previous subsection, this formula differs from \cite{Saad:2019lba} by a factor of $2 \sinh \frac{b}{2}$, which can be accounted for by comparing the measure of the boundary particle quantum mechanics and the measure on the trumpet wiggles used in \cite{Saad:2019lba}.\footnote{See footnote 79 in \cite{Jafferis:2022wez}.} Thus, the boundary particle formalism is able to compute the correct trumpet partition function.

\subsection{Disk}
\label{sec:disk}

We now review the disk computation performed in \cite{Yang:2018gdb}. We want to compute
\begin{equation}
	\frac{\int_{-\infty}^\infty  d\phi_1 \int_{-\infty}^\infty d\psi_1 d\psi_2 e^{- \psi_1 - \psi_2} \, \int_{\phi_1}^{\phi_1 + 2 \pi} d\phi_2	\,
		\calk_{\beta_2}(\phi_1 + 2 \pi ,\psi_1,v_1;\phi_2,\psi_2,v_2) \calk_{\beta_1}(\phi_2,\psi_2,v_2;\phi_1,\psi_1,v_1)}{\text{Vol } \widetilde{SL}(2,\mathbb{R})}.
	\label{eq:unfixeddisk}
\end{equation}
Note that the integrand does not actually depend on $v_1$ or $v_2$. To perform the Fadeev-Popov gauge-fixing procedure, we define
\begin{equation}
	\Delta(\phi_1,\phi_2,\psi_1,\psi_2) := 8\int dg \, \delta(\phi_1^g) \delta(\phi_2^g - \pi) \delta(\psi_1^g - \psi_2^g), \label{eq:Delta}
\end{equation}
where the $g$ integration is over the $\widetilde{SL}(2,\mathbb{R})$ manifold and $dg$ is the Haar measure. We wish to fix to a gauge where $\phi_1 = 0$, $\phi_2 = \pi$, and $\psi_1 = \psi_2$. Because the infinitesimal transformations of $\phi$ and $\psi$ in \eqref{eq:first}, \eqref{eq:second}, and \eqref{eq:third} do not depend on $\psi$, $\Delta(\phi_1,\phi_2,\psi_1,\psi_2)$ just evaluates to a constant. Because the overall normalization of the disk is ambiguous, we will simply insert \eqref{eq:Delta} into \eqref{eq:unfixeddisk}. The final result is
\begin{equation}
	8\int_{-\infty}^\infty d\psi \, e^{- 2\psi} 
	\calk_{\beta_2}(2 \pi ,\psi,0;\pi,\psi,0) \calk_{\beta_1}(\pi,\psi,0;0,\psi,0).
	\label{eq:2.39}
\end{equation}
For our later applications, suppose that we wanted to fix the renormalized length between points 1 and 2 in this computation. We simply need to insert a delta function into \eqref{eq:unfixeddisk}:
\begin{equation}
	\delta\left(\ell + 2 \log \left(\frac{ e^{\frac{\psi_1 + \psi_2}{2}}}{2\sin \left(\frac{\phi_2 - \phi_1}{2}\right)}\right) \right),
\end{equation}
where we refer to \eqref{eq:length} for the definition of renormalized length. With this insertion, \eqref{eq:2.39} becomes
\begin{align}
	8\int_{-\infty}^\infty d\psi \, e^{- 2\psi} 
	\calk_{\beta_2}(2 \pi ,\psi,0;\pi,\psi,0) \calk_{\beta_1}(\pi,\psi,0;0,\psi,0) 
\delta(\ell + 2 \psi - 2 \log 2)
= \Psi_{\beta_2}(\ell)\Psi_{\beta_1}(\ell),
\end{align}
where
\begin{align}
	\label{eq:2.42}
	\Psi_{\beta}(\ell) &:= \left. 2e^{-\psi}\calk_{\beta}(\pi,\psi,0;0,\psi,0)\right|_{\psi =  \log 2 - \frac{\ell}{2}}
	\\
	&=    \int_{0}^{\infty}ds \, \rho(s) e^{-\frac{s^2}{4 \phi_b}\beta} \, 2 K_{2is}\left(4 e^{-\ell/2}\right),
\end{align}
and the disk density of states $\rho(s)$ is defined by
\begin{equation}
	\rho(s) := \frac{s}{\pi^2} \sinh(2 \pi s).
\end{equation}
Noting that the Bessel functions obey
\begin{equation} \int_{-\infty}^\infty d\ell \, 2K_{2 i s }(4e^{-\ell/2}) \, 2K_{2 i s^\prime}(4e^{-\ell/2}) = \frac{\delta(s  - s^\prime)}{ \rho(s)},
\label{eq:bessel}
\end{equation}
we may readily compute the disk as follows:
\begin{equation}
	\int_{-\infty}^\infty  d\ell \, \Psi_{\beta_2}(\ell) \Psi_{\beta_1}(\ell).
\end{equation}
As was shown in \cite{Yang:2018gdb}, $\Psi_{\beta}(\ell)$ is the Hartle-Hawking wavefunction in the length basis. To recapitulate, the Hartle-Hawking wavefunction may be computed by slicing the disk into two half-disks. This calculation is summarized in Figure \ref{fig:disktwopoint}, but without the red dots (which represent matter operator insertions). We will generalize this computation to include matter in the following sections.

\section{Matter on the disk}
\label{sec:matterondisk}

In this section, we review aspects of the quantization of matter in AdS$_2$.

\begin{figure}
	\centering
	\includegraphics[width=0.5\linewidth]{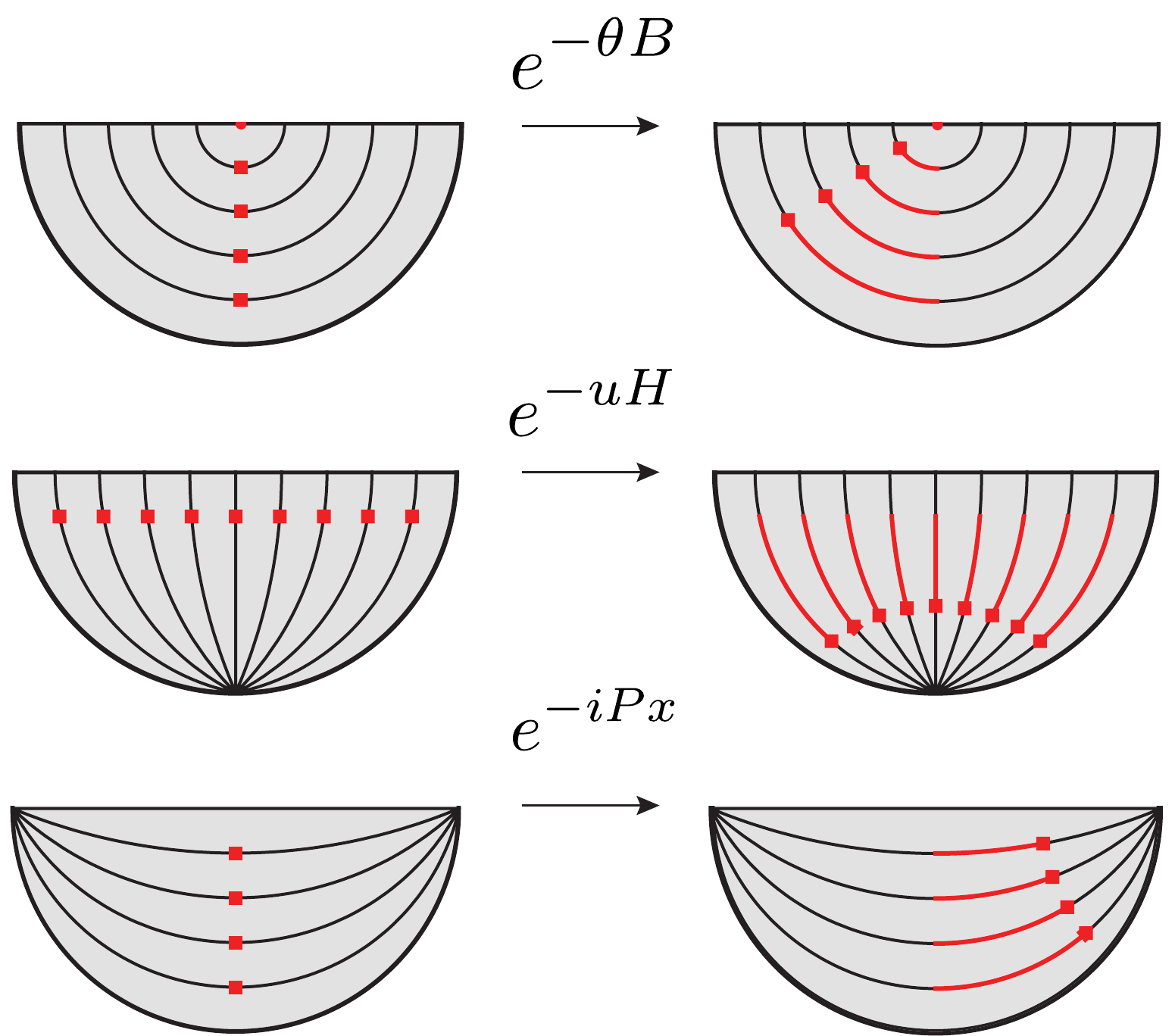}
	\caption{The Euclidean path integral prepares states in the Hilbert space $\calh_0$, which is associated to the horizontal interval boundary. The red dots on the left represent the locations where $\varphi$ is inserted in the lower-half disk. Under the action of a symmetry transformation, the matter operators move along the red lines to their final positions, as shown. Here, $\theta$, $u$, and $x$ are positive.}
	\label{fig:symmetryfigure}
\end{figure}

\pagebreak

A QFT quantized on global AdS$_2$ has three Hermitian spacetime symmetry generators: $B$, $H$, and $P$. A state in the Hilbert space $\calh_0$ may be prepared by inserting a local operator $\varphi$ somewhere in the Euclidean half-disk. To fix our conventions, we show in Figure \ref{fig:symmetryfigure} how the generators act on such states. Their algebra is given as follows:
\begin{equation}
	[B,H] = i P, \quad [H,P] = i B, \quad [B,P] = i H.
\label{eq:generators}
\end{equation}
A bulk operator $\varphi$, when extrapolated to the boundary, becomes a boundary operator $\bdyO$. For simplicity, let $\varphi$ be a scalar field. Working in polar coordinates,
\begin{equation}
	ds^2 = d\rho^2 + d\phi^2 \sinh^2 \rho,
\end{equation}
we define the boundary operator by
\begin{equation}
\label{eq:bdyop}
	\bdyO(\phi) := \lim_{\rho \rightarrow \infty} e^{\Delta \rho} \varphi(\phi,\rho).
\end{equation}
The parameter $\Delta$ depends on the specific operator and theory under consideration. We will assume that the matter theory is unitary, so $\Delta > 0$. We let $\ket{\bdyO(\phi)}$ for $\phi \in (0,\pi)$ refer to the state that is prepared by inserting \eqref{eq:bdyop} into the Euclidean path integral on the half-disk.

To see that \eqref{eq:generators} defines an $\mathfrak{sl}(2,\mathbb{R})$ algebra, we define
\begin{equation}
	L_0 := H, \quad L_{\pm 1} := B \pm i P,
\end{equation}
such that
\begin{equation}
	[L_n,L_m] = (n - m) L_{n + m}.
\end{equation}
All states in the matter QFT may be organized into primaries and descendants of this algebra. The matter Hilbert space is a direct sum of unitary lowest-weight discrete series representations and the trivial representation (which is the vacuum state). These are the unitary representations for which $H$ is bounded from below. There is a state-operator correspondence, where operators inserted at the bottom of the half-disk (that is, $\phi = \frac{\pi}{2}$ and $\rho = \infty$, see Figure \ref{fig:coordconventions}) correspond to states on the interval boundary of the half-disk. Hence, $\ket{\bdyO(\frac{\pi}{2})}$ is a primary, while $\ket{\bdyO(\phi)}$ for $\phi \neq \frac{\pi}{2}$ is a descendant.

We deduce the transformations of $\ket{\bdyO(\phi)}$ as follows:
\begin{align}
\label{eq:Htrans}
	e^{-u H} \ket{\bdyO(\phi)} &= \left( \frac{d \phi^{g_{u}}}{d \phi} \right)^{\Delta} \ket{\bdyO(\phi^{g_{u}})}, \quad {g_{u}} := e^{u T_3}, \\
	\label{eq:Ptrans}
	e^{- i P x} \ket{\bdyO(\phi)}  &= \left( \frac{d \phi^{g_{x}}}{d \phi} \right)^{\Delta} \ket{\bdyO(\phi^{g_{x}})}, \quad {g_{x}} := e^{-x T_2}, \\
	\label{eq:Btrans}
	e^{-\theta B} \ket{\bdyO(\phi)}  &= \left( \frac{d \phi^{g_{\theta}}}{d \phi} \right)^{\Delta} \ket{\bdyO(\phi^{g_{\theta}})}, \quad {g_{\theta}} := e^{- \theta T_1}.
\end{align}

The vacuum state, which is annihilated by all of the symmetry generators, will be denoted by $\ket{0}$.

Above, we showed how a bulk field $\varphi$ may be used to prepare a single primary state and its descendants, which constitute a discrete series representation. A general QFT has infinitely many primary states.  Each normalized primary state (other than the vacuum) will be denoted by $\ket{\bdyO_i(\frac{\pi}{2})}$, where $i$ indexes the choice of primary state. Then, we define $\ket{\bdyO_i(\phi)}$ for $\phi \in (0,\pi)$ by acting on $\ket{\bdyO_i(\frac{\pi}{2})}$ with the symmetry generators using \eqref{eq:Htrans}, \eqref{eq:Ptrans}, \eqref{eq:Btrans}. The state-operator correspondence asserts that $\ket{\bdyO_i(\phi)}$ may be prepared by inserting a primary operator $\bdyO_i(\phi)$ in the half-disk Euclidean path integral, localized in the asymptotic AdS region at angular coordinate $\phi$.\footnote{In other words, there is a bulk field for which an equation analogous to \eqref{eq:bdyop} holds for $\bdyO_i(\phi)$. The field could be, for example, a normal-ordered product of two $\varphi$ fields.}

The inner product of $\ket{\bdyO_i(\phi_1)}$ and $\ket{\bdyO_j(\phi_2)}$ may be computed using the Euclidean path integral on the disk. We write
\begin{equation}
\braket{\bdyO_j(\phi_2)| \bdyO_i(\phi_1)} = \braket{\overline{\bdyO_j}(\overline{\phi_2}) \bdyO_i(\phi_1)},
\end{equation}
where the bar over $\bdyO_j$ refers to an involution (or conjugation) on the space of primary operators and $\overline{\phi} := 2 \pi - \phi$ is the reflection of $\phi$ across the horizontal boundary of the half-disk. The involution $\bdyO_j \rightarrow \overline{\bdyO_j}$ does not change the scaling dimension. Without loss of generality, we assume that $\overline{\bdyO_j} = \bdyO_j$. Conformal invariance then implies that the two-point correlator, which is computed using the Euclidean path integral, obeys
\begin{equation}
\label{eq:mattertwopoint}
\braket{\bdyO_j(\phi_2) \bdyO_i(\phi_1)} = \frac{\delta_{ij}}{\left( \sin \frac{\phi_2 - \phi_1}{2} \right)^{2 \Delta_i}}.
\end{equation}
Because $\phi = 0$ and $\phi = \pi$ lie on the interval that is associated with $\calh_0$, we let $\bdyO_i(0)$ and $\bdyO_i(\pi)$ define local operators\footnote{Strictly speaking, local operators in QFT are actually operator-valued distributions. Later, we will address the question of how to construct bounded operators from local operators.} that act on $\calh_0$. The condition $\overline{\bdyO_i} = \bdyO_i$ implies that these operators are Hermitian. Their matrix elements define the structure constants:
\begin{equation}
C_{ijk} := \braket{ \bdyO_i\left(\frac{\pi}{2}\right) | 2^{-\Delta_j} \bdyO_j(\pi) | \bdyO_k\left(\frac{\pi}{2}\right) } = \braket{ \bdyO_i\left(\frac{\pi}{2}\right) | 2^{-\Delta_k} \bdyO_k(0) | \bdyO_j\left(\frac{\pi}{2}\right) },
\end{equation}
and conformal invariance implies that
\begin{equation}
	\label{eq:threept}
\braket{\bdyO_i(\phi_3)\bdyO_j(\phi_2)\bdyO_k(\phi_1)} =
\frac{C_{ijk}}{\left( \sin \frac{\phi_2 - \phi_1}{2} \right)^{\Delta_k + \Delta_j - \Delta_i} \left( \sin \frac{\phi_3 - \phi_1}{2} \right)^{\Delta_k + \Delta_i - \Delta_j} \left( \sin \frac{\phi_3 - \phi_2}{2} \right)^{\Delta_i + \Delta_j - \Delta_k}}.
\end{equation}
We take the structure constants $C_{ijk}$ to be real because they are computed using a Euclidean path integral. It follows that $C_{ijk}$ is invariant under arbitrary permutations of the indices.

To understand how matter operator insertions are treated in the boundary particle formalism, we use \eqref{eq:2.7} to evaluate a renormalized $\varphi(\phi,\rho)$ field at the position of the boundary particle at time $\tau$. We have that
\begin{equation}
	\lim_{\epsilon \rightarrow 0} \left(\frac{4 \phi_b}{\epsilon}\right)^\Delta \varphi(\phi(\tau),\rho(\tau)) = \lim_{\epsilon \rightarrow 0} \left(\frac{4 \phi_b}{\epsilon}\right)^\Delta \varphi(\phi(\tau),\log \frac{4 \phi_b}{\epsilon} - \psi(\tau)) = e^{\Delta \psi(\tau)}	\bdyO(\phi(\tau)).
\end{equation}

Note that $e^{\Delta \psi} \bdyO(\phi)$ is invariant under transformations that act on both $\bdyO$ and the $\phi,\psi$ coordinates,
\begin{equation}
	e^{\Delta \psi} \bdyO(\phi) = e^{u H}e^{\Delta \psi^{g_u}} \bdyO(\phi^{g_u}) e^{- u H} = e^{i P x} e^{\Delta \psi^{g_x}} \bdyO(\phi^{g_x}) e^{-i P x} = e^{\theta B}e^{\Delta \psi^{g_\theta}} \bdyO(\phi^{g_\theta}) e^{- \theta B}. 
\end{equation}

Thus, an insertion of $e^{\Delta \psi(\tau)} \bdyO(\phi(\tau))$ in the boundary particle path integral \eqref{eq:Euclideanschwarzianpathintegral} corresponds to an insertion of an operator on the boundary at time $\tau$.

\pagebreak

\section{Path integral quantization of JT with matter}

\label{sec:pathintegralquant}

We now describe the path integral quantization of JT with matter. In section \ref{sec:disk}, we explained how to slice a disk into two half-disks, with each half-disk preparing a Hartle-Hawking state. Likewise, in section \ref{sec:matterondisk}, we showed how matter states are prepared using the Euclidean path integral on the half-disk. We combine both calculations in this section. The result is equivalent to canonically quantized JT gravity with matter in Lorentzian signature, as discussed in \cite{Penington:2023dql}.

\begin{figure}
	\centering
	\includegraphics[width=\linewidth]{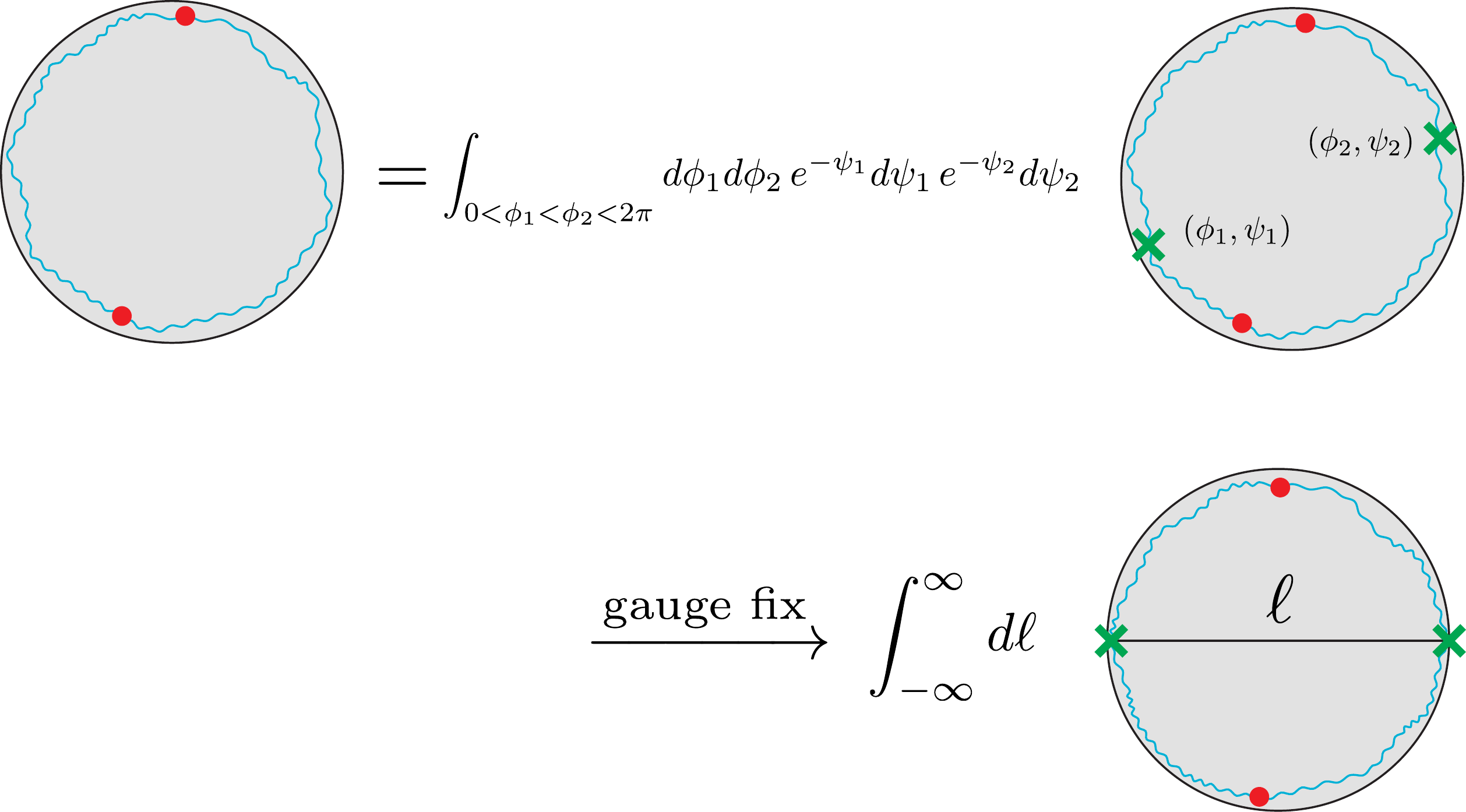}
	\caption{On the top-left, we depict the disk two-point function calculation, where operators (denoted by red dots) are inserted along the trajectory of the boundary particle at specified boundary times. To compute this two-point function, we pick two arbitrary boundary times (labeled by green crosses). The locations of the boundary particle at those times are given by $(\phi_1,\psi_1)$ and $(\phi_2,\psi_2)$. We explicitly integrate over these locations and use the boundary particle propagator to account for all of the possible trajectories (the integral is identical to \eqref{eq:unfixeddisk} but with matter operator insertions). The gauge-fixing condition requires that the green crosses are equidistant from the center of the disk and have angular coordinates $\phi = 0$ and $\phi = \pi$ as shown. After gauge-fixing, the four-dimensional integral is reduced to a single dimensional integral over the renormalized length $\ell$. We also sum over a complete set of states of the matter theory. Thus, an orthonormal basis state of the Hilbert space is labeled by $\ell$ together with an orthonormal basis state of the matter Hilbert space.}
	\label{fig:disktwopoint}
\end{figure}

See Figure \ref{fig:disktwopoint} for a depiction of the disk two-point function calculation. It is identical to the empty disk calculation of section \ref{sec:disk}, except we insert two additional operators as shown. Using the same gauge-fixing techniques as in section \ref{sec:disk}, we may write the two-point function as an overlap of two states, where each state is prepared using the Euclidean path integral with one matter operator insertion. The result is that the Hilbert space of JT with matter is
\begin{equation}
\label{eq:decomp}
	\calh := \calh_0 \otimes L^2(\mathbb{R}),
\end{equation}
where $L^2(\mathbb{R})$ represents square-integrable functions of the length mode $\ell$. The $\calh_0$ factor may be thought of as the Hilbert space of the matter theory on a Cauchy slice of renormalized length $\ell$. The $\mathfrak{sl}(2,\mathbb{R})$ generators acting on $\calh_0$ are the physical generators discussed in \cite{Lin:2019qwu}.

A Hartle-Hawking state with a single matter insertion is given by
\begin{equation}
	\begin{split}
	\ket{\Psi^{\bdyO}_{\beta_1,\beta_2}(\ell)} 
	:=	\int_{-\infty}^\infty d\psi_1 e^{- \psi_1} \int_0^\pi d\phi_1 \biggl[& 2e^{-\psi_0} \calk_{\beta_2}(\pi,\psi_0,0;\phi_1,\psi_1,v_1) \\
	&e^{\Delta \psi_1 } \ket{ \bdyO(\phi_1) } \,  \calk_{\beta_1}(\phi_1,\psi_1,v_1;0,\psi_0,0) \biggr|_{\psi_0 =  \log 2 - \frac{\ell}{2}} \biggr],
	\label{eq:HHmatter}
\end{split}
\end{equation}
which is the same as \eqref{eq:2.42} with the extra matter insertion in the path integral. Note that the $\psi$ variable is always integrated in the range $(-\infty,\infty)$.

\begin{figure}
	\centering
	\includegraphics[width=0.88\linewidth]{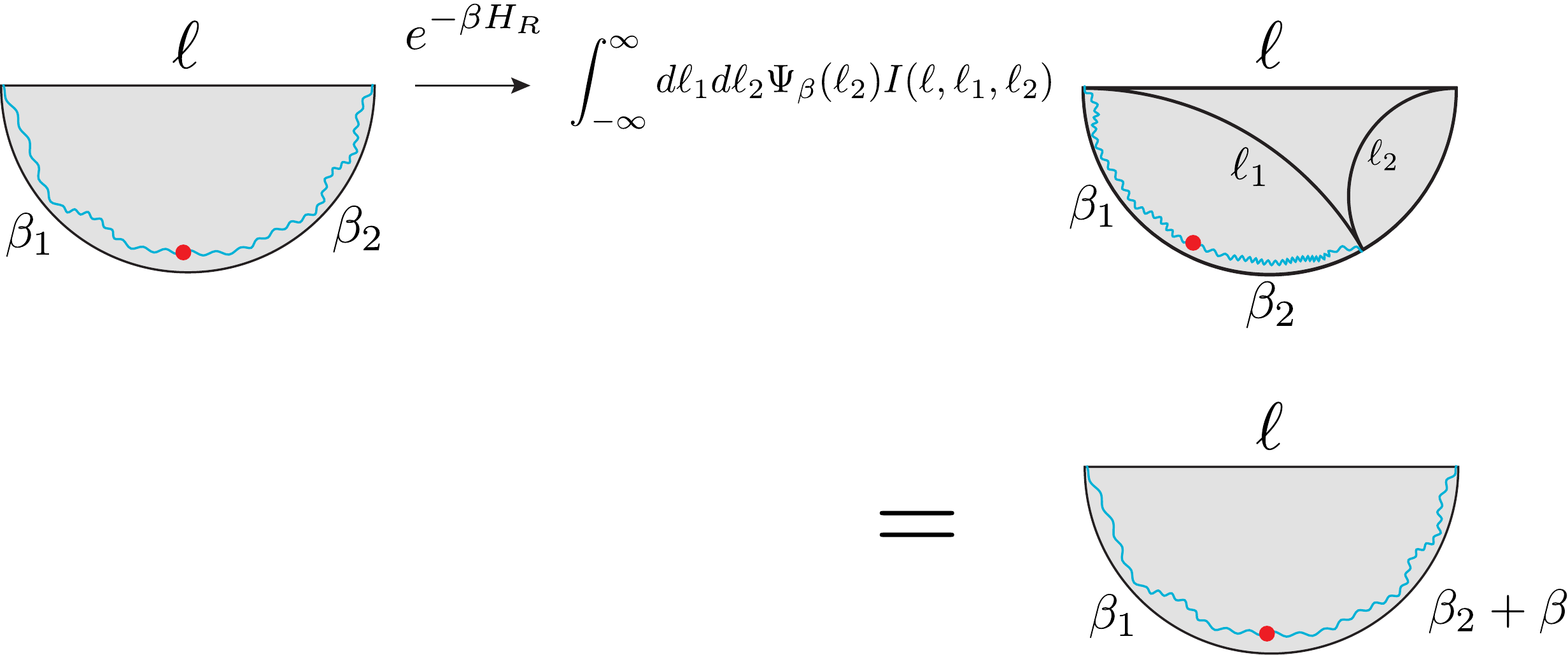}
	\caption{We pictorally represent equation \eqref{eq:consisehr}. The half-disk represents a state in the matter sector. The horizontal $\ell$ geodesic is centered on the origin. On the left, we prepare a state using the JT+matter Euclidean path integral. The red dot represents an insertion of a renormalized matter field along the trajectory of the boundary particle. The effect of $e^{-\beta H_R}$ is to adjoin an additional boundary particle trajectory of renormalized length $\beta$ to the original one (and integrate over all the possible ways of doing this). The lengths $\ell_1$ and $\ell_2$ determine (via trilateration) the location where the wiggly blue trajectory of length $\beta_2$ terminates. The $\Psi_\beta(\ell_2) I(\ell,\ell_1,\ell_2)$ factor computes the amplitude for the boundary particle to propagate from this location to a position in the upper-right corner along a trajectory of length $\beta$. By integrating over $\ell_1$ and $\ell_2$, we integrate over the location where the $\beta_2$ and $\beta$ trajectories are attached.}
	\label{fig:ehrfig}
\end{figure}

We now compute the right- and left- Hamiltonians $H_R$ and $H_L$. In the absence of matter, we have $H_R = H_L$, but they differ in general. The reader may choose to skip this derivation. The final equations are boxed and are summarized in Figure \ref{fig:ehrfig}. We now focus on computing $H_R$. By definition, we have that
\begin{equation}
	\begin{split}
		\label{eq:4.3}
		e^{-\beta H_R} \ket{\Psi^{\bdyO}_{\beta_1,\beta_2}(\ell)} &:= \ket{\Psi^{\bdyO}_{\beta_1,\beta_2 + \beta}(\ell)} \\
		&= \int_{-\infty}^\infty d\psi_1 e^{- \psi_1} \,  d\psi e^{- \psi} \int_{0 < \phi_1 < \phi < \pi} d\phi_1  d\phi \,  \biggl[ 2e^{-\psi_0}
		\\ 
		&\quad  \calk_{\beta}(\pi,\psi_0,0;\phi,\psi,v)
		\calk_{\beta_2}(\phi,\psi,v;\phi_1,\psi_1,v_1) 
		\\
		&\quad   e^{\Delta \psi_1 } \ket{ \bdyO(\phi_1) } \,  \calk_{\beta_1}(\phi_1,\psi_1,v_1;0,\psi_0,0) \biggr|_{\psi_0 =  \log 2 - \frac{\ell}{2}}  \biggr].
	\end{split}
\end{equation}
Next, we seek a transformation that moves the geodesic connecting $(0,\psi_0)$ and $(\phi,\psi)$ to a horizontal geodesic centered at the origin of our polar coordinate system, which by definition extends from $(0,\psi_2)$ to $(\pi,\psi_2)$. In particular, define $\tilde{g}$ and $\psi_2$ such that
\begin{equation}
(\pi^{\tilde{g}},\psi_2^{\tilde{g}}) = (\phi,\psi), \quad \text{ and } \quad (0^{\tilde{g}},\psi_2^{\tilde{g}}) = (0,\psi_0).
\end{equation}
Note that $\tilde{g}$ and $\psi_2$ depend on the integration variables $\phi,\psi$. Explicitly, we have
\begin{align}
e^{\psi_2} &:= \frac{e^{\frac{\psi + \psi_0}{2}}}{\sin \frac{\phi}{2}}, \label{eq:psi2} \\
\tilde{g} &:= e^{ -  \cot \left( \frac{\phi}{2} \right)  (T_1 - T_3)} e^{(\psi_2 - \psi_0) T_2}.
\end{align}
Then, \eqref{eq:4.3} becomes
\begin{equation}
	\begin{split}
		&\int_{-\infty}^\infty d\psi_1 e^{- \psi_1} \,  d\psi e^{- \psi} \int_{0 < \phi_1 < \phi < \pi} d\phi_1  d\phi \, \biggl[ 2e^{-\psi_0}
		\\ 
		&\quad \quad \calk_{\beta}(\pi,\psi_0,0;\phi,\psi,v^{\tilde{g}})
		\calk_{\beta_2}(\pi^{\tilde{g}},\psi_2^{\tilde{g}},v^{\tilde{g}};\phi_1,\psi_1,v_1) 
		\\
		&\quad \quad  e^{\Delta \psi_1 } \ket{ \bdyO(\phi_1) } \, \calk_{\beta_1}(\phi_1,\psi_1,v_1;0^{\tilde{g}},\psi_2^{\tilde{g}},0^{\tilde{g}})e^{0^{\tilde{g}}} \biggr|_{\psi_0 =  \log 2 - \frac{\ell}{2}}  \biggr].
	\end{split}
	\label{eq:4.5}
\end{equation}
We can replace $v$ with $v^{\tilde{g}}$ because the integrand does not actually depend on $v$. Likewise, we replace the last argument of $\calk_{\beta_1}$ with a $0^{\tilde{g}}$ and include a compensating $e^{0^{\tilde{g}}}$ term. Explicitly, we have that
\begin{align}
	\label{eq:4.8}
e^{0^{\tilde{g}}} &= e^{-e^{\psi_0} \cot \left( \frac{\phi}{2}\right)},
\\
\label{eq:4.9}
v^{\tilde{g}} &= v + e^{\psi} \cot \left(\frac{\phi}{2}\right).
\end{align}
Next, we define a symmetry transformation $\tilde{V}$ that acts on $\calh_0$ such that
\begin{equation}
	\label{eq:4.10}
	e^{\Delta \psi_1^{\tilde{g}}} \bdyO(\phi_1^{\tilde{g}}) = \tilde{V} e^{\Delta \psi_1} \bdyO(\phi_1) \tilde{V}^{-1}.
\end{equation}
Explicitly, we have that
\begin{equation}
\tilde{V} := e^{- \cot \left(\frac{\phi}{2}\right) (B + H)} e^{-i P(\psi_0 - \psi_2)}.
\label{eq:tildeV}
\end{equation}
We then perform a $\tilde{g}$ transformation on the integration variables $\phi_1,\psi_1$ in \eqref{eq:4.5} and use \eqref{eq:4.10} together with the invariance of $\calk_{\beta_2}$ and $\calk_{\beta_1}$ to obtain
\begin{equation}
	\begin{split}
		&\int_{-\infty}^\infty d\psi_1 e^{- \psi_1} \,  d\psi e^{- \psi} \int_{0}^\pi d\phi \int_{0}^\pi d\phi_1   \,  \biggl[ 2e^{-\psi_0} 
		\\ 
		&\quad \quad \calk_{\beta}(\pi,\psi_0,0;\phi,\psi,v^{\tilde{g}})
		\calk_{\beta_2}(\pi,\psi_2,v;\phi_1,\psi_1,v_1) 
		\\
		&\quad \quad  \tilde{V} e^{\Delta \psi_1 } \ket{ \bdyO(\phi_1) } \,  \calk_{\beta_1}(\phi_1,\psi_1,v_1;0,\psi_2,0)e^{0^{\tilde{g}}} \biggr|_{\psi_0 =  \log 2 - \frac{\ell}{2}} \biggr].
	\end{split}
	\label{eq:4.7}
\end{equation}
Using \eqref{eq:HHmatter}, \eqref{eq:4.8}, and \eqref{eq:4.9}, we may rewrite \eqref{eq:4.7}  as follows:
\begin{equation}
	\begin{split}
		e^{-\beta H_R} \ket{\Psi^{\bdyO}_{\beta_1,\beta_2}(\ell)}  &= \int_{-\infty}^\infty d\psi \, e^{-\psi} \int_0^\pi d\phi \, \biggl[
		\calk_{\beta}(\pi,\psi_0,0;\phi,\psi,0)
		\\
		&\quad \frac{e^{\frac{\psi - \psi_0}{2}}}{\sin \frac{\phi}{2}} e^{-(e^{\psi} + e^{\psi_0}) \cot \left(\frac{\phi}{2}\right)}
		\tilde{V} \ket{\Psi^{\bdyO}_{\beta_1,\beta_2}\left(2 \log 2 - 2 \psi_2\right)} \biggr|_{\psi_0 =  \log 2 - \frac{\ell}{2}} \biggr].
	\end{split}
\label{eq:4.13}
\end{equation}
We now perform a change of variables from $\phi$,$\psi$ to $\ell_1$,$\ell_2$ ,
\begin{equation}
	e^{-\ell_1/2} := \frac{e^{\frac{\psi_0 + \psi}{2}}}{2 \sin \frac{\phi}{2}}, \quad e^{-\ell_2/2} := \frac{ e^{ \frac{\psi_0 + \psi}{2}} }{2 \cos \frac{\phi}{2}}.
\end{equation}
and we obtain
\begin{equation}
	\begin{split}
		e^{-\beta H_R} \ket{\Psi^{\bdyO}_{\beta_1,\beta_2}(\ell)}  &= \int_{-\infty}^\infty d\ell_1 d\ell_2 \int_0^\infty ds \, \biggl[ \frac{s \sinh (2 \pi s)}{\pi^2} e^{- \frac{s^2}{4 \phi_b} \beta} K_{2is}(4 e^{-\ell_2/2})  
		\\
		&e^{-2 e^{-\frac{1}{2}(\ell + \ell_1 + \ell_2)} (e^{\ell} + e^{\ell_1} + e^{\ell_2})} e^{- e^{\frac{\ell_2 - \ell_1}{2}} (H+B)} e^{- i P \frac{\ell_1 - \ell}{2}} \ket{\Psi^\bdyO_{\beta_1,\beta_2}(\ell_1)}\biggr].
	\end{split}
\end{equation}
\noindent
We may write this more concisely as follows:
\begin{equation}
	\begin{split}
\boxed{		e^{-\beta H_R} \ket{\Psi^{\bdyO}_{\beta_1,\beta_2}(\ell)}  = \int_{-\infty}^\infty d\ell_1 d\ell_2 \Psi_\beta(\ell_2) I(\ell,\ell_1,\ell_2) e^{- e^{\frac{\ell_2 - \ell_1}{2}} (H+B)} e^{- i P \frac{\ell_1 - \ell}{2}} \ket{\Psi^\bdyO_{\beta_1,\beta_2}(\ell_1)},}
	\end{split}
\label{eq:consisehr}
\end{equation}
where
\begin{equation}
	\begin{split}
	I(\ell,\ell_1,\ell_2) &:= \frac{1}{2}e^{-2 e^{-\frac{1}{2}(\ell + \ell_1 + \ell_2)} (e^{\ell} + e^{\ell_1} + e^{\ell_2})} 
	\\
	&= \int_0^\infty ds \, \rho(s) \, 2 K_{2 i s}(4 e^{-\ell/2}) 2 K_{2 i s}(4 e^{-\ell_1/2}) 2 K_{2 i s}(4 e^{-\ell_2/2}).
	\end{split}
\label{eq:Idef}
\end{equation}

The analogous expression for $H_L$ is
\begin{equation}
	\begin{split}
\boxed{		e^{-\beta H_L} \ket{\Psi^{\bdyO}_{\beta_1,\beta_2}(\ell)}  = \int_{-\infty}^\infty d\ell_1 d\ell_2 \Psi_\beta(\ell_2) I(\ell,\ell_1,\ell_2) e^{- e^{\frac{\ell_2 - \ell_1}{2}} (H-B)} e^{ i P \frac{\ell_1 - \ell}{2}} \ket{\Psi^\bdyO_{\beta_1,\beta_2}(\ell_1)}.}
	\end{split}
\label{eq:concisehl}
\end{equation}
To obtain an explicit expression for $H_R$, note that \eqref{eq:Euclideanschwarzianpathintegral} implies that
\begin{align}
	\left.- \frac{d}{d\beta}K_\beta(\phi_2,\psi_2,\phi_1,\psi_1)\right|_{\beta = 0} &= \braket{\phi_2,\psi_2|H|\phi_1,\psi_1} 
	\\
	&= \frac{1}{2 \phi_b} \left[ -\frac{1}{2}\partial_{\psi_2}^2 + e^{\psi_2} \partial_{\phi_2} - \frac{1}{2}e^{2 \psi_2} \right] \delta(\phi_2 - \phi_1) \delta(\psi_2 - \psi_1).
\end{align}
Starting from \eqref{eq:4.13} and differentiating with respect to $\beta$ at $\beta = 0$, we obtain delta functions that we use to perform the $d\phi$ $d\psi$ integration. After further simplifications, we finally obtain
\begin{equation}
	H_R \ket{\Psi^{\bdyO}_{\beta_1,\beta_2}(\ell)} = \frac{1}{2 \phi_b} \left[- \frac{1}{2} \partial^2_{\ell} + \frac{i P}{2} \partial_\ell + \frac{P^2}{8} + e^{-\ell/2}(B + H) + 2 e^{-\ell} \right] \ket{\Psi^{\bdyO}_{\beta_1,\beta_2}(\ell)},
\end{equation}
from which it follows that\footnote{To reach this conclusion, we need to show that states of the form $\ket{\Psi^{\bdyO_i}_{\beta_1,\beta_2}(\ell)}$, for $\bdyO_i$ ranging over all matter primaries, span the Hilbert space. We do this in appendix \ref{sec:spanning}.}
\begin{equation}
\boxed{	2 \phi_b H_R = \frac{1}{2}\left[k - \frac{P}{2}\right]^2 + (H + B) e^{-\ell/2} + 2 e^{-\ell},}
\end{equation}
where $k$ is the conjugate operator to $\ell$ that obeys $[\ell,k] = i$. A similar computation for $H_L$ tells us that
\begin{equation}
\boxed{	2 \phi_b H_L = \frac{1}{2}\left[k + \frac{P}{2}\right]^2 + (H - B) e^{-\ell/2} + 2 e^{-\ell}.}
\end{equation}
Using \eqref{eq:generators}, it is straightforward to check that the expressions for $H_R$ and $H_L$ above commute. Furthermore, $H_R$ and $H_L$ are positive operators.\footnote{We prove that $H+B$ and $H-B$ are nonnegative operators in appendix \ref{sec:positive}.}

As mentioned earlier, to include matter, we insert a renormalized $\varphi$ field at the location of the boundary particle. In the gauge-fixed description, the boundary particle is located at the end of the $\ell$ geodesic, or a renormalized distance of $\ell/2$ away from the center of our polar coordinate system. Thus, we define $\calo_R$ by inserting a matter operator on the right boundary in the gauge-fixed description.
\begin{equation}
\label{eq:caloRdef}
	\calo_R := \lim_{\epsilon \rightarrow 0} \left(\frac{4 \phi_b}{\epsilon}\right)^\Delta \varphi(\pi,\ell/2 + \log \frac{2 \phi_b}{\epsilon}) = \left(2e^{-\ell/2}\right)^{\Delta} \bdyO(\pi).
\end{equation}
Likewise, $\calo_L$ is defined by
\begin{equation}
\label{eq:caloLdef}
	\calo_L :=
	\lim_{\epsilon \rightarrow 0} \left(\frac{4 \phi_b}{\epsilon}\right)^\Delta \varphi(0,\ell/2 + \log \frac{2 \phi_b}{\epsilon}) = \left(2e^{-\ell/2}\right)^{\Delta} \bdyO(0).
\end{equation}
It is straightforward to check that $[\calo_L,H_R] = [\calo_R,H_L] = 0$. To allow us to place operators at other locations on the geodesic, we define
\begin{equation}
	\bdyO_R(w) := (2e^{-w})^\Delta \bdyO(\pi), \quad \quad 	\bdyO_L(w) := (2e^{-w})^\Delta \bdyO(0). 
	\label{eq:orol}
\end{equation}
These operators are located a renormalized distance $w$ away from the center of the geodesic.

The definitions \eqref{eq:caloRdef} and \eqref{eq:caloLdef} define boundary operators dual to the bulk field $\varphi$. For a general primary operator $\bdyO_i$, one may define the operators $\calo_{i,R}$ and $\calo_{i,L}$ in terms of $\bdyO_i(\pi)$ and $\bdyO_i(0)$ using the rightmost expressions in \eqref{eq:caloRdef} and \eqref{eq:caloLdef}. In general, we will not write the $i$ index unless we need to refer to multiple primary operators in the same context.

\begin{figure}
	\centering
	\includegraphics[width=0.7\linewidth]{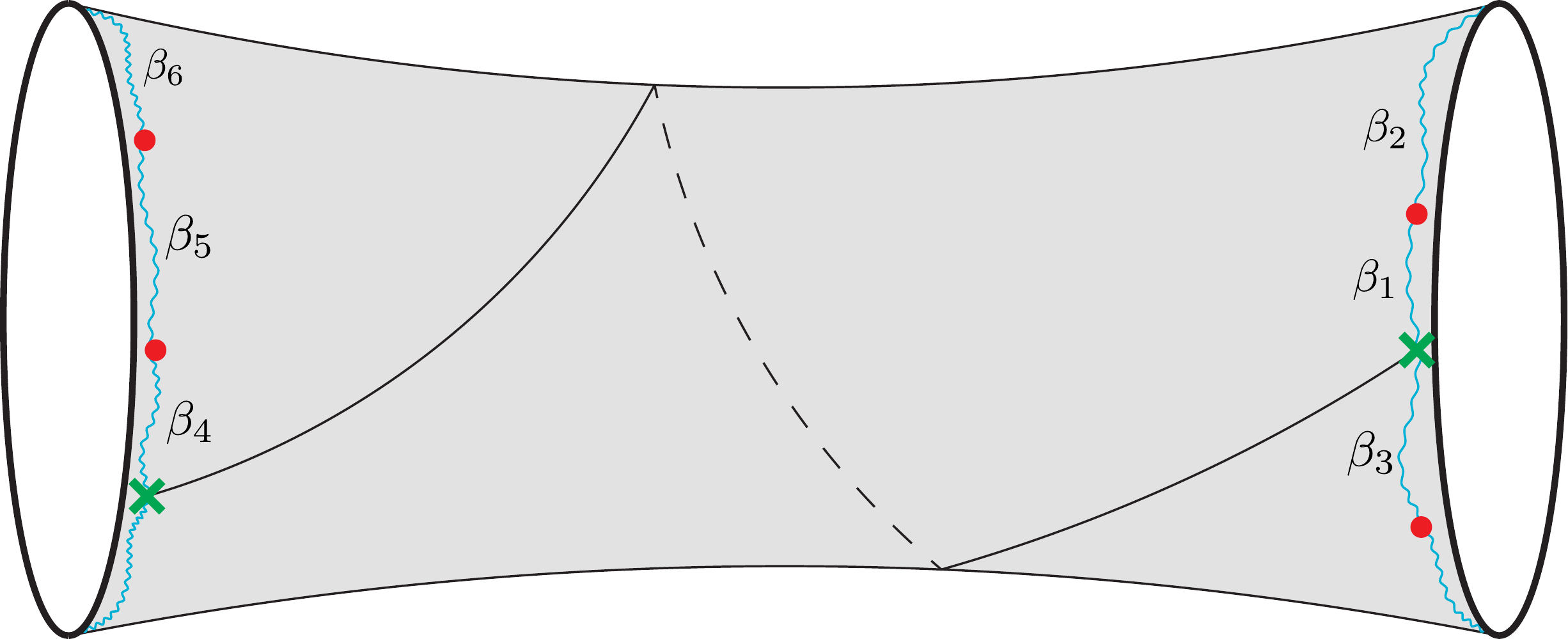}
	\caption{We depict the path integral that computes \eqref{eq:bigtrace}. The red dots represent operator insertions. The green crosses mark the $\tau = 0$ points on each boundary. We sum over all geodesics that connect the two $\tau = 0$ points. We show one such geodesic here.}
	\label{fig:doubletrumpetfig}
\end{figure}

At this point, we may compute more complicated expressions such as
\begin{equation}
	\text{Tr } e^{-\dt \ell} e^{-\beta_6 H_L} \calo_L e^{-\beta_5 H_L} \calo_L e^{-\beta_4 H_L} e^{-\beta_3 H_R} \calo_R e^{-\beta_2 H_R} \calo_R e^{-\beta_1 H_R},
\label{eq:bigtrace}
\end{equation}
where $\dt > 0$ is unrelated to $\Delta$. See appendix \ref{sec:appendixa} for the detailed computation. The result is simply the double-trumpet path integral of JT with matter, with a sum over all geodesics that connect the $\tau = 0$ points on either boundary, weighted by $e^{-\dt \ell}$. We illustrate this path integral in Figure \ref{fig:doubletrumpetfig}. Without the matter, this result is already known \cite{Saad:2019pqd,Blommaert:2020seb,Lin:2022zxd,Jafferis:2022wez}. Here, we have shown that the result generalizes straightforwardly when matter is included. Note that this path integral is divergent because the integral over double-trumpet geometries includes geometries where the closed geodesic that wraps the neck of the double-trumpet becomes arbitrarily small. This corresponds to an arbitrarily high temperature for the matter fields and leads to a universal divergence. To avoid this divergence, one would have to insert a highly non-local operator into \eqref{eq:bigtrace}, such as a projection onto the vacuum state in the matter sector. We will refer to this divergence again in section \ref{sec:factorization}.

\subsection{The energy basis}
\label{sec:energybasis}

Having introduced $H_R$, $H_L$, $\calo_R$, $\calo_L$, and $\ell$, we now find explicit expressions for the matrix elements of these operators in the energy basis. As shown in appendix \ref{sec:spanning}, there exists a complete set of orthonormal states that are labeled by a primary operator in the matter sector and the eigenvalues of $H_R$ and $H_L$. We refer to this basis as the energy basis. Thus, we may re-express $\calh$ as follows:
\begin{equation}
\calh = L^2(\mathbb{R}^+) \oplus \bigoplus_{\cals_p} \left[  L^2(\mathbb{R}^+) \otimes L^2(\mathbb{R}^+) \right].
\end{equation}
The index set $\cals_p$ is defined to be the set of matter primary operators other than the identity. The first $L^2(\mathbb{R}^+)$ refers to the subspace of $\calh$ obtained by acting on $\calh$ with the projection $\ket{0}\bra{0}  \otimes 1 $, where we are referring to the decomposition in \eqref{eq:decomp}. This subspace is spanned by square-integrable functions of the energy, which is positive. A basis state in this subspace is denoted by $\ket{0; s}$, where $s^2 /(4 \phi_b)$ is the (right or left) energy and $s > 0$. We refer to this subspace as the vacuum sector of $\calh$. Each of the remaining terms in the direct sum refers to the subspace of $\calh$ obtained by acting with the projection $P_i \otimes 1$, where $P_i$ is the projection onto a single discrete series representation within $\calh_0$ that corresponds to the primary operator $\bdyO_i \in \cals_p$. A state in this subspace is given by a square-integrable function of the left and right energies. A basis state in this subspace is denoted by $\ket{i ; s_L, s_R}$ where $s_L$ and $s_R$ are valued in $\mathbb{R}^+$, the positive real line. We refer to this subspace as a primary sector of $\calh$.

We work in conventions where a projection onto the vacuum sector is given by
\begin{equation}
\label{eq:res1}
\int_0^\infty \rho(s) \, \ket{0 ; s}\bra{0 ; s},
\end{equation}
and a projection onto one of the primary sectors labeled by $i$ is given by
\begin{equation}
\label{eq:res2}
\int_0^\infty ds_L \, \rho(s_L) \, d s_R \, \rho(s_R) \, \ket{i ; s_L \, s_R} \bra{i ; s_L \, s_R}.
\end{equation}
The presence of the disk density of states in \eqref{eq:res1} and \eqref{eq:res2} is purely a matter of convention, since the $s$ variables are continuous.

We denote the Hartle-Hawking state, which lies in the vacuum sector, as
\begin{equation}
\ket{\beta} := \int_{-\infty}^\infty d\ell \, \Psi_\beta(\ell) \,  \ket{0} \otimes \ket{\ell} ,
\end{equation}
where $\Psi_\beta(\ell)$ was defined in \eqref{eq:2.42}, and $\ket{\ell}$ is an eigenstate of $\ell$, normalized as
\begin{equation}
\braket{\ell_1 | \ell_2} = \delta(\ell_1 - \ell_2).
\end{equation}
We refer to the state prepared by the half-disk Euclidean path integral with an operator insertion as
\begin{equation}
\ket{ i ; \beta_L , \beta_R} := \int_{-\infty}^\infty d\ell \,  \ket{\Psi^{\bdyO_i}_{\beta_L,\beta_R} (\ell)} \otimes \ket{\ell} ,
\end{equation}
where $\ket{\Psi^{\bdyO_i}_{\beta_L,\beta_R} (\ell)}$ was defined in \eqref{eq:HHmatter}.

Using
\begin{equation}
\braket{\beta_1 | \beta_2} = \int_0^\infty ds \, \rho(s) \, e^{-\frac{s^2}{4 \phi_b} (\beta_1 + \beta_2)} = \int_0^\infty ds \, \braket{\beta_1 | 0;s} \rho(s) \braket{0;s | \beta_2},
\end{equation}
we have
\begin{equation}
\braket{ 0 ; s | \beta } = e^{-\frac{s^2}{4 \phi_b} \beta}.
\end{equation}
Next, we consider
\begin{align}
\braket{j ; \beta_L^\prime , \beta_R^\prime | i ; \beta_L , \beta_R} &= \sum_{k \in \cals_p} \int_0^\infty ds_L \, ds_R \, \braket{j ; \beta_L^\prime , \beta_R^\prime | k ; s_L, s_R} \rho(s_L) \rho(s_R) \braket{ k ; s_L, s_R | i ; \beta_L , \beta_R}
\\
\label{eq:twopoint2}
&= \braket{ \beta_L + \beta_L^\prime | \calo_{i,L} \calo_{j,R} | \beta_R + \beta_R^\prime }
\\
&=\delta_{ij} \int_0^\infty ds_L \, ds_R \, \rho(s_L) \, \rho(s_R) \, \Gamma^{\Delta_i}_{s_L, s_R} e^{- \frac{s_L^2}{4 \phi_b} (\beta_L + \beta_L^\prime)} e^{- \frac{s_L^2}{4 \phi_b} (\beta_R + \beta_R^\prime)},
\end{align}
where we define
\begin{equation}
\Gamma^{\Delta}_{s_1,s_2} := \int_{-\infty}^\infty d\ell \, 2 K_{2 i s_1}(4 e^{-\ell/2}) \, 4^\Delta e^{-\Delta \ell} \, 2 K_{2 i s_2}(4 e^{-\ell/2}) = \prod_{\pm \pm} \frac{\Gamma(\Delta \pm i s_1 \pm i s_2)}{\Gamma(2 \Delta)}, 
\label{eq:besselidentity}
\end{equation}
and the product is over all four choices of signs. Equation \eqref{eq:twopoint2} is the two-point function of $\bdyO_i$ and $\bdyO_j$ on the disk, integrated over all metrics using the boundary particle propagator. We may use \eqref{eq:mattertwopoint}, \eqref{eq:length}, and \eqref{eq:2.42} to explicitly write \eqref{eq:twopoint2} as an integral over the renormalized length $\ell$ between the the two points where the operators are inserted. In particular, the two-point function is computed by integrating over all of the boundary particle trajectories with an additional weighting of $4^\Delta e^{- \Delta \ell}$. It follows that
\begin{equation}
\braket{j ; s_L, s_R | i ; \beta_L, \beta_R} = \delta_{ij} \sqrt{\Gamma^{\Delta_i}_{s_L, s_R}}e^{- \frac{s_L^2}{4 \phi_b} \beta_L} e^{- \frac{s_L^2}{4 \phi_b} \beta_R}.
\end{equation}

We now consider matrix elements of $\calo_{i,R}$ in the energy basis. Because one-point functions of primary operators other than the identity vanish, we have that
\begin{equation}
\braket{ 0 ; s_2 | \calo_{i,R} | 0 ; s_1 } = 0.
\end{equation}

\begin{figure}[!h]
	\centering
	\includegraphics[width=0.7\linewidth]{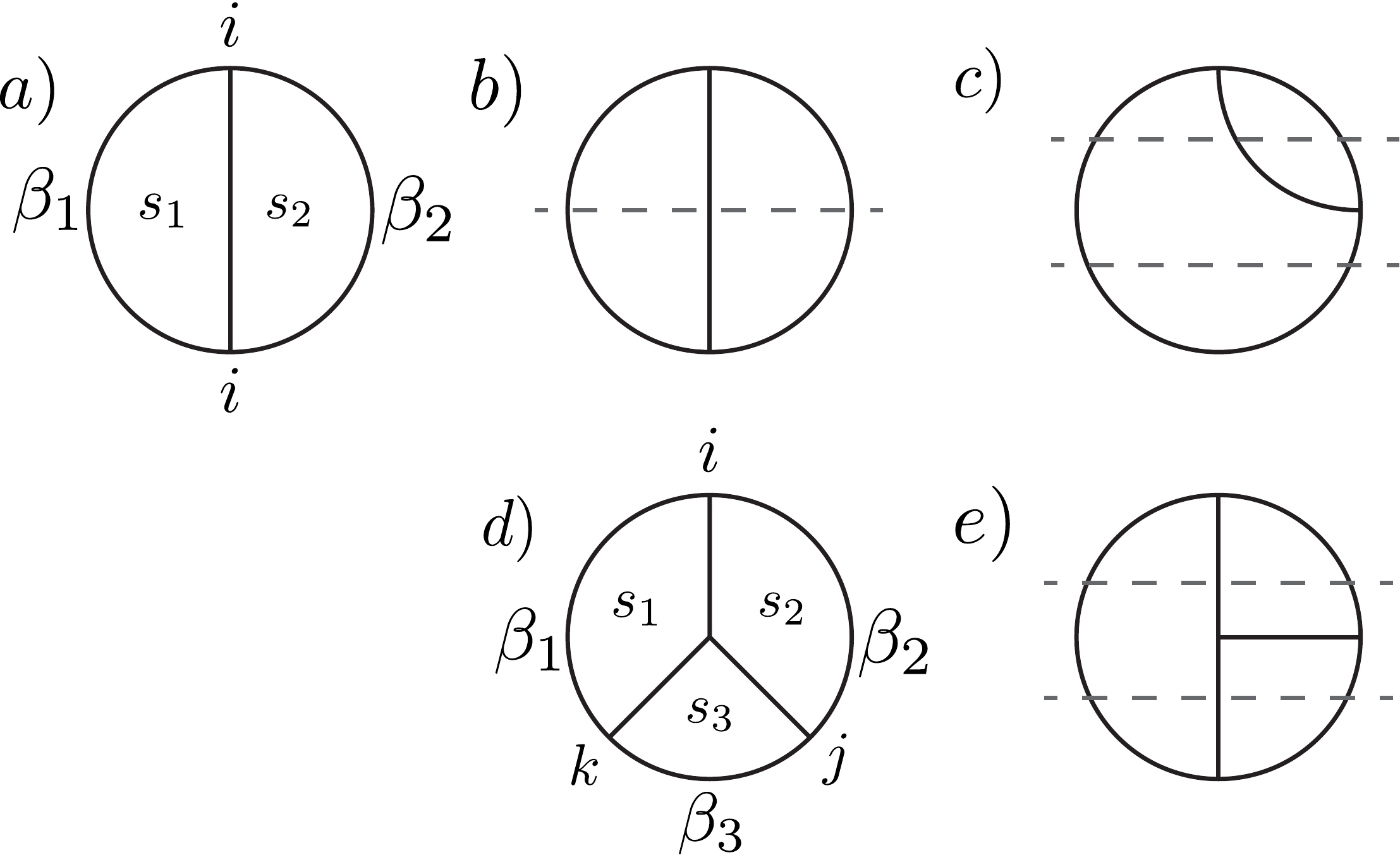}
	\caption{We illustrate how to rewrite disk correlators as matrix elements of $\calo_R$ in the energy basis. a) The diagrammatic representation of the disk two-point function, which corresponds to \eqref{eq:twopoint}. b) We may view the disk two-point function as an overlap between a bra and a ket, where the bra and ket wavefunctions may be computed by separately applying the diagrammatic rules described in the main text to the regions above and below the dashed line. c) The dashed lines separate the diagram into three regions. The top and bottom regions prepare bra and ket wavefunctions. The middle region is interpreted as a matrix element of $\calo_R$, namely \eqref{eq:4.43}. d) The disk three-point function may be written as \eqref{eq:threepoint}, and the corresponding diagram includes a bulk trivalent vertex, as shown. e) We separate the three-point function diagram into three regions. Again, the top and bottom regions prepare states, while the middle region is a matrix element, see \eqref{eq:threepointmatrixelem}. To summarize, we may use the disk two- and three-point functions to compute the matrix elements of $\calo_R$ and $\calo_L$ in the energy basis. Using these matrix elements, we may compute any disk correlator with arbitrary matter insertions. The resulting correlators are given by the diagrammatic rules described in the main text.}
	\label{fig:diagrams}
\end{figure}

\pagebreak

Next, we use
\begin{equation}
 \braket{ \beta_L | \calo_{i,R} e^{- \beta_R H_R} \calo_{i,R} | \beta} = \int_0^\infty ds_L \, ds_R \, ds \, \braket{ i ; \beta_L, \beta_R | i ; s_L, s_R} \rho(s_L) \rho(s_R) \braket{ i ; s_L , s_R | \calo_{i,R} | 0 ; s } \rho(s) \braket{0 ; s | \beta} 
\end{equation}
to infer that
\begin{equation}
\braket{ j ; s_L , s_R | \calo_{i,R} | 0 ; s } = \delta_{ij} \frac{\delta(s_L - s)}{\rho(s)} \sqrt{\Gamma^{\Delta_i}_{s_L, s_R}}.
\label{eq:4.43}
\end{equation}

The calculations in this subsection may be conveniently interpreted using a set of diagrammatic rules. These diagrammatic rules are similar to the rules described in \cite{Mertens:2017mtv,Jafferis:2022wez}. See the diagrams in Figure \ref{fig:diagrams}. Every closed region is assigned an $s$ parameter, or an energy. Each diagram represents an integral over the $s$ parameters with measure $\rho(s) \, ds$. For every three-way junction of the black lines, the integrand contains a function of the neighboring $s$ parameters. A black line on the circumference is assigned an inverse temperature, and the integrand includes a corresponding Boltzmann factor. A bulk black line extending from the circumference, which we label with an $i$ index, represents the insertion of a primary $\bdyO_i$, and is associated with the function $\sqrt{\Gamma^{\Delta_i}_{s_1,s_2}}$, where $s_1$, $s_2$ are the neighboring $s$ parameters. When three bulk black lines meet in the interior, we include the function $V_{ijk}(s_1,s_2,s_3)$, which will be determined below. For example, diagram a) in Figure \ref{fig:diagrams}, which represents the disk two-point function of two primaries, is given by
\begin{equation}
\int_0^\infty ds_1 \, \rho(s_1) \, ds_2 \, \rho(s_2) \, \Gamma^{\Delta_i}_{s_1,s_2} \, e^{- \frac{s_1^2}{4 \phi_b} \beta_1} \, e^{- \frac{s_2^2}{4 \phi_b} \beta_2},
\label{eq:twopoint}
\end{equation}
while diagram d), which represents a disk three-point function of primaries $\bdyO_i$, $\bdyO_j$, and $\bdyO_k$, is given by
\begin{equation}
\label{eq:threepoint}
\int_0^\infty ds_1 \, \rho(s_1) \, ds_2 \, \rho(s_2) \, ds_3 \, \rho(s_3) \, \sqrt{\Gamma^{\Delta_i}_{s_1,s_2} \Gamma^{\Delta_j}_{s_2,s_3} \Gamma^{\Delta_k}_{s_3,s_1} } \, e^{- \frac{s_1^2}{4 \phi_b} \beta_1} \, e^{- \frac{s_2^2}{4 \phi_b} \beta_2} \, e^{- \frac{s_3^2}{4 \phi_b} \beta_3} \, V_{ijk}(s_1,s_2,s_3).
\end{equation}
Finally, using the strategy described in Figure \ref{fig:diagrams} e), we compute
\begin{equation}
	\braket{k ; s_L^\prime, s_R^\prime | \calo_{j,R} | i ; s_L, s_R} = \frac{\delta(s_L - s_L^\prime)}{\rho(s_L)} V_{ijk}(s_L,s_R,s_R^\prime)\sqrt{\Gamma^{\Delta_i}_{s_L^\prime,s_R^\prime}}.
	\label{eq:threepointmatrixelem}
\end{equation}
To compute $V_{ijk}(s_1,s_2,s_3)$, we need to compute the three-point function and show that it takes the form \eqref{eq:threepoint}. We insert the QFT three-point function \eqref{eq:threept} into the disk path integral. The additional weighting factor we insert is
\begin{equation}
	C_{ijk} \left(2 e^{-\ell_{ij}/2}\right)^{\Delta_i + \Delta_j - \Delta_k} \left(2 e^{-\ell_{ki}/2}\right)^{\Delta_i + \Delta_k - \Delta_j} \left(2 e^{-\ell_{jk}/2}\right)^{\Delta_j + \Delta_k - \Delta_i},
\end{equation}
where $\ell_{ij}$, $\ell_{jk}$, and $\ell_{ki}$ are the renormalized geodesic distances between the insertions of $\bdyO_i$, $\bdyO_j$, and $\bdyO_k$. The three-point function becomes
\begin{equation}
	\begin{split}
	&\int_{-\infty}^\infty d\ell_{ij} \, d\ell_{jk} \, d\ell_{ki} \, \Psi_{\beta_2}(\ell_{ij}) \Psi_{\beta_3}(\ell_{jk}) \Psi_{\beta_1}(\ell_{ki}) I(\ell_{ij},\ell_{jk},\ell_{ki})
	\\
	&\quad \times C_{ijk} \left(2 e^{-\ell_{ij}/2}\right)^{\Delta_i + \Delta_j - \Delta_k} \left(2 e^{-\ell_{ki}/2}\right)^{\Delta_i + \Delta_k - \Delta_j} \left(2 e^{-\ell_{jk}/2}\right)^{\Delta_j + \Delta_k - \Delta_i}, 
\end{split}
\label{eq:threepointprimaries}
\end{equation}
where $I(\ell_{ij},\ell_{jk},\ell_{ki})$ was defined in \eqref{eq:Idef}. Because we want to show that \eqref{eq:threepointprimaries} is equivalent to \eqref{eq:threepoint}, we use \eqref{eq:2.42} to take the inverse Laplace transform of \eqref{eq:threepointprimaries} with respect to $\beta_1$, $\beta_2$, and $\beta_3$. We obtain
\begin{equation}
	\begin{split}
	& \sqrt{\Gamma^{\Delta_i}_{s_1,s_2} \Gamma^{\Delta_j}_{s_2,s_3} \Gamma^{\Delta_k}_{s_3,s_1} } \,  V_{ijk}(s_1,s_2,s_3)
	\\
		&= \int_{-\infty}^\infty d\ell_{ij} \, d\ell_{jk} \, d\ell_{ki} \, \, 
		2K_{2 i s_2}(4e^{-\ell_{ij}/2})
		2K_{2 i s_3}(4e^{-\ell_{jk}/2})
		2K_{2 i s_1}(4e^{-\ell_{ki}/2})
	    I(\ell_{ij},\ell_{jk},\ell_{ki})
	\\
	&\quad \times C_{ijk} \left(2 e^{-\ell_{ij}/2}\right)^{\Delta_i + \Delta_j - \Delta_k} \left(2 e^{-\ell_{ki}/2}\right)^{\Delta_i + \Delta_k - \Delta_j} \left(2 e^{-\ell_{jk}/2}\right)^{\Delta_j + \Delta_k - \Delta_i}.
	\end{split}
\label{eq:4.49}
\end{equation}
We are only interested in evaluating this integral for positive values of all of the scaling dimensions. In appendix \ref{sec:threepointapp}, we simplify the integral and show that $V_{ijk}(s_1,s_2,s_3)$ is well-defined. The explicit expression for $V_{ijk}(s_i,s_j,s_k)$ will not be important for us, but we note that it is invariant under permutations of $i$, $j$, and $k$.

Finally, we consider matrix elements of $\ell$ in the energy basis. Because these are not needed to prove our main results in section \ref{sec:algebras}, we will be brief. It is easiest to consider the matrix elements of $e^{- \dt \ell}$ for $\dt > 0$, which are
 \begin{align}
  \braket{0 ;s^\prime  |  e^{- \dt \ell} | 0; s } &= 4^{-\dt} \Gamma^{\dt}_{s,s^\prime}, \\
 \braket{j ;s_L^\prime \, s_R^\prime |  e^{- \dt \ell} | i; s_L \, s_R } &=  \delta_{ij} \, 4^{-\dt} \left\{\begin{array}{ccc}
			\dt & s_L & s_L^\prime \\
			\Delta & s_R^\prime & s_R
		\end{array}
		\right\} \sqrt{\Gamma^{\dt}_{s_L,s_L^\prime}  \Gamma^{\dt}_{s_R,s_R^\prime} }.
		\label{eq:6j}
 \end{align}
The special function in \eqref{eq:6j} is a 6j symbol of $\mathfrak{sl}(2,\mathbb{R})$, which is defined in \cite{Jafferis:2022wez}.

\subsection{Conformal blocks and the OPE}

In this subsection, we show how the energy basis matrix elements allow us to produce compact expressions for conformal blocks. We also discuss the OPE in the presence of gravity.

\begin{figure}
	\centering
	\includegraphics[width=\linewidth]{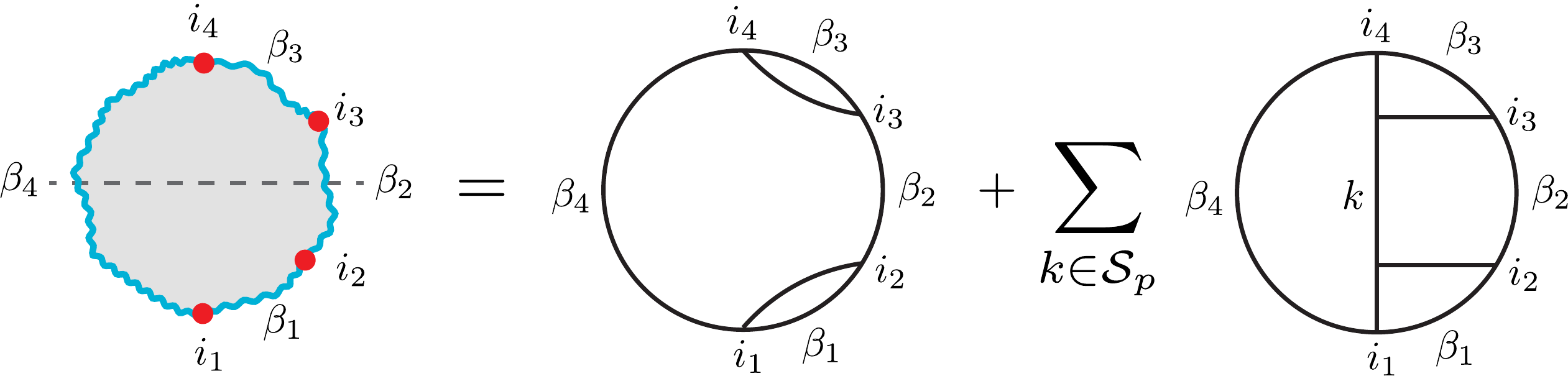}
	\caption{We depict the disk correlator of four primary operators. The dashed line indicates where we insert a complete set of states in the energy basis. In the first term on the right, the exchanged states lie in the vacuum sector. In the second term, the exchanged states are in the primary sectors. The diagrams on the right side may be evaluated using the rules described in section \ref{sec:energybasis}. Alternatively, we could have drawn the dashed line to pass through $\beta_3$ and $\beta_1$ instead of $\beta_2$ and $\beta_4$. These two options correspond to the $s$- and $t$-channel decompositions of the four-point function into conformal blocks. In particular, a single term in the sum on the right hand side equals a four-point conformal block coupled to JT gravity.}
	\label{fig:fourptfig}
\end{figure}

To start, consider the four-point function of arbitrary primaries. It is given by
\begin{equation}
\label{eq:4pt}
\braket{ i_4 ; \epsilon, \beta_3 | \calo_{i_3,R} e^{-\beta_2 H_R} \calo_{i_2,R} | i_1 ; \beta_4 - \epsilon, \beta_1 },
\end{equation}
which is independent of the parameter $\epsilon$ that obeys $0 < \epsilon < \beta_4$. By inserting a resolution of the identity in the energy basis and using the matrix elements computed in the previous subsection, we find that \eqref{eq:4pt} is equal to
\begin{align}
&\delta_{i_1 i_2} \delta_{i_3 i_4}  \int_0^\infty ds_1 \rho(s_1) \, ds_2 \rho(s_2) \, ds_3 \rho(s_3) \,  \Gamma^{\Delta_{i_1}}_{s_1, s_2} \Gamma^{\Delta_{i_3}}_{s_2, s_3} e^{- \frac{s_1^2}{4 \phi_b} \beta_1} e^{- \frac{s_2^2}{4 \phi_b} (\beta_2 + \beta_4)} e^{- \frac{s_3^2}{4 \phi_b} \beta_3}
\\
&+ \sum_{k \in \cals_p} \int_0^\infty \prod_{j = 1}^4 \left[ ds_j \rho(s_j) e^{- \frac{s_j^2}{4 \phi_b} \beta_j} \sqrt{\Gamma^{\Delta_{i_j}}_{s_{j},s_{j+1}}}  \right] V_{i_1 i_2 k}(s_4,s_1,s_2) V_{k i_3 i_4}(s_4,s_2,s_3),
\end{align}
where $s_5 := s_1$. The calculation is summarized in Figure \ref{fig:fourptfig}. In the semiclassical limit, the four-point function may be decomposed into $SL(2,\mathbb{R})$ conformal blocks. Away from the semiclassical limit, it is still possible to compute the contribution from a single exchanged primary. Each diagram on the right side of Figure \ref{fig:fourptfig} refers to such a contribution. The generalization to arbitrary $n$-point diagrams is clear: an $n$-point conformal block corresponds to a tree graph with trivalent vertices drawn in the interior of a circle, with $n$ bulk lines attached to the circumference. An explicit expression in the energy basis may be obtained by applying the diagrammatic rules described in section \ref{sec:energybasis}.\footnote{An explicit check of these claims is performed in appendix C of \cite{Jafferis:2022wez}, where the matter theory is taken to be a free scalar field.}

One may generalize the OPE of two primary operators away from the semiclassical limit. By working in the energy basis, one may verify that the following OPE identity holds:
\begin{equation}
\begin{split}
\label{eq:4.52}
e^{-\epsilon H_R} \calo_{i,R} e^{-\beta H_R} \calo_{j,R} e^{-\epsilon H_R} &= \sum_{n \ge 0}  a_n H_R^n e^{-2 \epsilon H_R} 
\\
&+ \sum_{k \in \cals_p} \sum_{n,m \ge 0} b^{(k)}_{n,m} H_R^n e^{- \epsilon H_R} \calo_{k,R} H_R^m e^{- \epsilon H_R} ,
\end{split}
\end{equation}
where $a_n$ and $b_{n,m}$ are Taylor coefficients defined from the following expansions,
\begin{align}
\delta_{ij} \int_0^\infty ds \, \rho(s) \Gamma^{\Delta_i}_{\sqrt{4 \phi_b E},s} \, e^{- \frac{s^2}{4 \phi_b} \beta}  &:= \sum_{n \ge 0} a_n E^n,
\\
\int_0^\infty ds \, \rho(s) \, V_{ijk}(\sqrt{4 \phi_b E_L},s,\sqrt{4 \phi_b E_R}) \sqrt{ \frac{\Gamma^{\Delta_i}_{\sqrt{4 \phi_b E_L},s} \Gamma^{\Delta_j}_{s, \sqrt{4 \phi_b E_R}}}{\Gamma^{\Delta_k}_{\sqrt{4 \phi_b E_L},\sqrt{4 \phi_b E_R}}}} e^{-\frac{s^2}{4 \phi_b} \beta} &:= \sum_{n,m \ge 0} b^{(k)}_{n,m} E_L^n E_R^m.
\end{align}
The functions $V_{ijk}(s_1,s_2,s_3)$ and $\Gamma^\Delta_{s_1,s_2}$ are even functions of the $s$ parameters, so only nonnegative integer powers appear in the Taylor expansions on the right side. One may insert either side of \eqref{eq:4.52} into a correlator and use the diagrammatic rules to obtain the same result.

\section{Algebras of boundary observables}

\label{sec:algebras}

Having reviewed various aspects of JT gravity with matter, our aim in this section is to define the von Neumann (vN) algebras that describe the physics accessible to observers who live on the AdS boundaries. A vN algebra is a $*$-algebra of bounded operators\footnote{The algebra of bounded operators on $\calh$ is denoted by $\calb(\calh)$.} that is equal to its double-commutant. In QFT, a physically interesting algebra is the algebra of all local operators belonging to a causally complete spacetime subregion. Because the fundamental fields are operator-valued distributions instead of operators, one needs to employ a smearing prescription to produce bounded operators from the fundamental fields. This is because the fundamental fields exhibit a universal UV divergence when they are brought close together. In our case, the fundamental QFT operators of interest are $\bdyO(\pi)$ and $\bdyO(0)$, because they are defined on the Cauchy slice associated with $\calh_0$. In JT gravity, these operators become $\calo_R$ and $\calo_L$, as reviewed earlier. Since the semiclassical limit $\phi_b \rightarrow \infty$ coincides with the UV limit, any prescription that produces bounded operators in QFT will suffice for producing bounded operators in JT gravity with matter. We define the set $\cals_R$ as follows,
\begin{equation}
\label{eq:boundedops}
\cals_R := \left\{ H_R^n e^{-\beta H_R} \, | \,  \beta > 0 , \, n \in \mathbb{Z}_{\ge 0} \right\} \cup \left\{ H_R^n e^{-\epsilon H_R} \calo_{i,R} e^{-\epsilon H_R} H_R^m \, | \, \epsilon > 0, \, i \in \cals_p, \, n,m \in \mathbb{Z}_{\ge 0} \right\},
\end{equation}
where $\cals_p$ is the set of primary operators in the matter QFT aside from the identity. All of the operators in $\cals_R$ are bounded. We define the algebra $\cala_{R,0}$ to be the algebra of finite linear combinations of finite products of operators in $\cals_R$. This is the algebra generated by words made from the letters $H_R$ and $\calo_{i,R}$, with Boltzmann factors inserted where necessary to keep the operators bounded. We have deliberately omitted the identity operator from $\cala_{R,0}$ for reasons that will become clear in section \ref{sec:tracesection}. The algebra $\cala_{R,0}$ is a $*$-algebra of bounded operators, but it is not equal to its double-commutant. We define the vN algebra $\cala_R$ to be the algebra generated by $\cala_{R,0}$,
\begin{equation}
\cala_R := \cala_{R,0}^{\prime \prime},
\end{equation}
where $^\prime$ denotes commutant. According to the bicommutant theorem \cite{jones}, $\cala_R$ is closed in the weak operator topology. In particular, an equivalent way to generate $\cala_R$ from $\cala_{R,0}$ is to include the limit points of all nets\footnote{For a review of nets and operator topologies, see \cite{Sorce:2023fdx}. Note that a sequence is an example of a net.}  $\{ a_\alpha \}$  that obey
\begin{equation}
\braket{x | a_\alpha | y} \rightarrow \braket{x | a | y}, \quad \forall  \ket{x}, \ket{y} \in \calh,
\end{equation}
where $\rightarrow$ denotes net convergence, $a_\alpha \in \cala_{R,0}$, and $a \in \calb(\calh)$. Note that the identity operator is the weak limit of $e^{-\beta H_R}$ as $\beta \rightarrow 0$. The algebra $\cala_R$ contains the right Hamiltonian and all of the $\calo_{i,R}$ operators, which in the context of AdS/CFT should be interpreted as the algebra of light (single-trace and multi-trace) operators of the right CFT. The left algebras $\cala_{L,0}$ and $\cala_{L}$ are defined analogously to $\cala_{R,0}$ and $\cala_{R}$. The OPE identity \eqref{eq:4.52}  furnishes an example of a weakly convergent sequence because it holds when inserted between any bra and ket.

Next, we prove that $\cala_R$ and $\cala_L$ are commutants of each other. Then, we will prove that these algebras are factors (namely, that $\cala_L \cap \cala_R = \mathbb{C} \cdot 1$). If the algebras were type I, these properties would imply that the Hilbert space $\calh$ admits a tensor product decomposition $\calh = \calh_L \otimes \calh_R$ where each algebra acts on one of the factors. Because the Hilbert space does not admit such a decomposition, it follows that $\cala_R$ and $\cala_L$ are not type I. In the next section, we will show that these algebras admit a (faithful, semifinite, and normal) trace that obeys $\text{tr } 1 = \infty$. This precludes the algebras from being type III or type II$_1$. It follows from the type classification of vN algebras \cite{Sorce:2023fdx} that $\cala_R$ and $\cala_L$ are type II$_\infty$ factors.

To prove that $\cala_R$ is the commutant of $\cala_L$, we let $R \in \calb(\calh)$ be a bounded operator that commutes with $H_L$ and $\calo_{i,L}$ for $i \in \cals_p$. We will then show that $R \in \cala_R$. We assume without loss of generality that $R$ is Hermitian. We make the following ansatz for the matrix elements of $R$ in the energy basis:
\begin{align}
\label{eq:Rfirst}
\braket{j ; s_L^\prime, s_R^\prime | R | i ; s_L, s_R } &= \frac{\delta(s_L^\prime - s_L)}{\rho(s_L)} R^{(1)}_{ji}(s_L,s_R,s_R^\prime),
\\
\label{eq:Rsecond}
\braket{j ; s_L^\prime, s_R^\prime | R | 0 ; s } &= \frac{\delta(s_L^\prime - s)}{\rho(s)} R^{(2)}_j(s,s_R^\prime),
\\
\label{eq:Rthird}
\braket{0 ; s^\prime | R | 0 ; s } &= \frac{\delta(s^\prime - s)}{\rho(s)} R^{(3)}(s).
\end{align}
Equations \eqref{eq:Rfirst}, \eqref{eq:Rsecond}, and \eqref{eq:Rthird} represent the most general form that these matrix elements can take given that $R$ commutes with $H_L$. Now we consider
\begin{equation}
\braket{j ; s_L, s_R | R \calo_{i,L} | 0 ; s} = \braket{j ; s_L, s_R | \calo_{i,L} R  | 0 ; s},
\end{equation}
which becomes, after inserting a resolution of the identity in the energy basis,
\begin{align}
 R_{ji}^{(1)}(s_L,s,s_R) \sqrt{\Gamma^{\Delta_i}_{s,s_L}}  &= \delta_{ij} \sqrt{\Gamma^{\Delta_i}_{s_L,s_R}}   \frac{\delta(s_R - s)}{\rho(s)} R^{(3)}(s) 
 +  \sum_{k \in \cals_p} V_{ijk}(s, s_L,s_R) \sqrt{\Gamma^{\Delta_i}_{s_L,s}}  R_k^{(2)}(s,s_R).
\end{align}
Dividing by $\sqrt{\Gamma^{\Delta_i}_{s,s_L}}$, we obtain
\begin{align}
 R_{ji}^{(1)}(s_L,s,s_R)   &= \delta_{ij}  \frac{\delta(s_R - s)}{\rho(s)} R^{(3)}(s) 
 +  \sum_{k \in \cals_p} V_{ijk}(s, s_L,s_R) R_k^{(2)}(s,s_R).
 \label{eq:eqforr1}
\end{align}
Equation \eqref{eq:eqforr1} fixes $R^{(1)}_{ji}$ given $R^{(2)}_k$ and $R^{(3)}$. Because the $s$ parameters are valued in $(0,\infty)$, we may assume that $R^{(2)}_k$ and $R^{(3)}$ are invariant under multiplying any of their arguments by $-1$ (in other words, only even powers of their arguments appear in a series expansion around zero). Equation \eqref{eq:eqforr1} implies the following series representation for $R$,
\begin{equation}
R = R^{(3)}(\sqrt{4 \phi_b H_R}) + \sum_{k \in \cals_p} \sum_{n,m \ge 0} r^{(2),k}_{n,m} H_R^n \calo_{k,R} H_R^m,
\label{eq:seriesexpansion}
\end{equation}
where $r^{(2),k}_{n,m}$ are the Taylor coefficients defined by
\begin{equation}
\sum_{n,m \ge 0} r^{(2),k}_{n,m} E_1^n E_2^m = \frac{1}{\sqrt{\Gamma^{\Delta_k}_{\sqrt{4 \phi_b E_1},\sqrt{4 \phi_b E_2}}}} R^{(2)}_k (\sqrt{4 \phi_b E_2}, \sqrt{4 \phi_b E_1}).
\end{equation}
One may verify \eqref{eq:seriesexpansion} in the energy basis. It follows that $R$ is in the weak closure of $\cala_{R,0}$, meaning that $R \in \cala_R$, or that $\cala_L^\prime \subset \cala_R$. It is straightforward to see that $\cala_R \subset \cala_L^\prime$ given that the generators of $\cala_R$ commute with the generators of $\cala_L$, which was verified in section \ref{sec:pathintegralquant}. Hence, $\cala_R = \cala_L^\prime$. Taking the commutant, we obtain $\cala_R^\prime = \cala_L$.

To prove that $\cala_R$ and $\cala_L$ are factors, we let $C \in \calb(\calh)$ be a bounded operator that lies in $\cala_L^\prime \cap \cala_R^\prime$. We again assume without loss of generality that $C$ is Hermitian. We need to show that $C$ is proportional to the identity operator. First, consider how $C$ acts on the vacuum sector. Let $\ket{\Psi}$ be an arbitrary normalized state in the vacuum sector. In particular, $\ket{\Psi}$ obeys
\begin{equation}
(H_L - H_R) \ket{\Psi} = 0.
\end{equation}
The wavefunction
\begin{equation}
(C\Psi)_i(s_L,s_R) := \braket{ i ; s_L, s_R | C | \Psi}
\end{equation}
must be a square-integrable function of $s_L$ and $s_R$ because $C$ is bounded. However, because $C$ commutes with $H_L - H_R$, it must be the case that $(C\Psi)_i(s_L,s_R)$ vanishes for $s_L \neq s_R$. Because its support has measure zero, this wavefunction has zero norm. Hence, $C$ maps the vacuum sector into the vacuum sector. Because $C$ is Hermitian, the image of $C$ acting on a primary sector is orthogonal to the vacuum sector. We can thus make the following ansatz for the non-zero matrix elements of $C$,
\begin{align}
\braket{j ; s_L^\prime, s_R^\prime | C | i ; s_L, s_R} &:= \frac{\delta(s_L - s_L^\prime)}{\rho(s_L)} \frac{\delta(s_R - s_R^\prime)}{\rho(s_R)} C^{(1)}_{ij}(s_L,s_R),
\\
\braket{0 ; s^\prime | C | 0 ; s} &:= \frac{\delta(s - s^\prime)}{\rho(s)} C^{(2)}(s).
\end{align}
Then, we consider
\begin{equation}
\label{eq:5.16}
\braket{j; s_L, s_R | C \calo_{i,L} | 0 ; s} = \braket{j; s_L, s_R |  \calo_{i,L} C | 0 ; s},
\end{equation}
which becomes, after inserting a complete set of states,
\begin{equation}
 \frac{\delta(s_R - s)}{\rho(s)} C^{(1)}_{ij}(s_L,s)  \sqrt{\Gamma^{\Delta_i}_{s,s_L}} = \delta_{ij} \sqrt{\Gamma^{\Delta_i}_{s,s_L}}  \frac{\delta(s_R - s)}{\rho(s)} C^{(2)}(s),
\end{equation}
which simplifies to
\begin{equation}
\label{eq:Ceq1}
  C^{(1)}_{ij}(s_L,s)   = \delta_{ij} C^{(2)}(s),
\end{equation}
and an analogous argument starting from \eqref{eq:5.16} but with $\calo_{i,L}$ replaced with $\calo_{i,R}$ implies that
\begin{equation}
  C^{(1)}_{ij}(s,s_R)   = \delta_{ij} C^{(2)}(s).
  \label{eq:Ceq2}
\end{equation}
Equations \eqref{eq:Ceq1} and \eqref{eq:Ceq2} together imply that $C_{ij}^{(1)}$ is a constant, independent of its arguments, which further implies that $C$ is proportional to the identity operator, which is what we wanted to prove.

A consequence of our results is that the union of $\cala_R$ and $\cala_L$ generates all of $\calb(\calh)$. More generally, if $\cala$ is a vN algebra, we have that
\begin{equation}(\cala \cap \cala^\prime)^\prime = (\cala \cup \cala^\prime)^{\prime \prime}.
\end{equation}
See \cite{Sorce:2023fdx} and proposition 1 in I.1 of \cite{19811}.\footnote{We thank Jon Sorce for pointing this formula out to us.} Setting $\cala = \cala_R$, we obtain
\begin{equation}
	(\cala_R\cup \cala_L)^{\prime \prime} = (\mathbb{C}\cdot 1)^\prime = \calb(\calh).
\end{equation}
In particular, bounded functions of the length operator are generated from the boundary algebras. In \cite{Lin:2022rbf}, the length operator was connected to the size operator in a pair of double-scaled SYK models. The length operator and the matter operators were all defined in terms of the fundamental Majorana fermions. In the limit where the theory becomes JT gravity with matter, we find that there must be a nontrivial relation directly between the length and matter operators.

\section{Traces}

\label{sec:tracesection}

In the previous section, we proved that $\cala_L$ and $\cala_R$ are type II$_\infty$ factors, which implies that they admit a unique (faithful, semifinite, and normal) trace up to an overall rescaling. Let $\text{tr}$ denote this trace (our normalization convention is specified in \eqref{eq:trace}). As was pointed out in \cite{Penington:2023dql}, this leads to a prescription for computing holographic entanglement entropy in the bulk Lorentzian theory. In the semiclassical limit, one may use the Quantum Extremal Surface (QES) formula, which separates the entropy into an area term and a bulk entropy term. In JT gravity with matter, the holographic entropy formula which resums all perturbative $G_N$ corrections\footnote{When $S_0$ is finite, there should be additional corrections.} is
\begin{equation}
\label{eq:heeformula}
S(\rho) = S_0 - \text{tr } \rho \log \rho,
\end{equation}
where $\rho \in \cala_R$ is a density matrix operator that defines a state (that is, $\rho$ is positive and normalized such that $\text{tr } \rho = 1$).\footnote{A normal state $\omega$ on $\cala_R$ (i.e. an abstractly defined positive ultraweakly continuous linear functional on the algebra) can be written as $\omega(\cdot) = \lim_n \text{tr } (\rho_n \cdot)$ where $\rho_n$ is a sequence of trace-class operators in $\cala_R$. See the end of section 6.1 of \cite{Sorce:2023fdx} and references therein.} The $S_0$ parameter reflects an ambiguity in the entropy that stems from the ambiguity in normalizing the trace. It can be fixed by matching to the entropy computed in the boundary dual theory.\footnote{An example of a boundary dual theory is the SYK model in the limit $N \gg 1$, $N/(\beta J)$ fixed.} Despite its appearance, \eqref{eq:heeformula} does not give a state-counting interpretation of the entropy, because $S_0$ does not correspond to an actual ground state degeneracy and because $\cala_R$ is not type I. Equation \eqref{eq:heeformula} should be viewed as a generalization of the QES formula away from the semiclassical limit, and its form is highly constrained by the fact that $\cala_R$ is a factor.

The formula for $\text{tr}$ was given in \cite{Penington:2023dql}. It is
\begin{equation}
\text{tr } a = \lim_{\beta \rightarrow 0} \braket{\beta | a | \beta}, \quad a \in \cala_R.
\label{eq:trace}
\end{equation}
Note that $\text{tr } 1 = \infty$. Equation \eqref{eq:trace} has all of the properties of a trace (which we list below). For $a \in \cala_{R,0}$, \eqref{eq:trace} is computable from the disk Euclidean path integral using the diagrammatic rules described in section \ref{sec:energybasis}, and the cyclicity of the trace is manifest in this description. The partition function formula $Z(\beta) = \text{tr } e^{-\beta H_R}$ implies a $\sinh$ density of states for $H_R$, given by $\rho(s)$. This $\sinh$ density of states was originally discovered from a Euclidean path integral calculation (such as in \cite{Stanford:2017thb}). The fact that the Lorentzian theory also knows about the $\sinh$ density is not necessarily surprising. In particular, in pure JT gravity, the $\sinh$ density of states was found from an analysis of the QES formula away from the semiclassical limit \cite{Jafferis:2019wkd}. A key difference between pure JT gravity and JT gravity with matter is that the $\sinh$ density is the only density permitted by the structure of the boundary algebras, whereas more information about the bulk theory is needed to argue that pure JT gravity knows about the $\sinh$.

In the remainder of this section, we emphasize the properties that the trace is required to have, and show how other traces may be defined once certain properties are discarded. In particular, the inevitability of the $\sinh$ density of states rests on certain assumptions about the trace that we will now state explicitly.

Per the theory of operator algebras \cite{Takesaki1979}, a (faithful, semifinite, normal) trace on the von Neumann algebra $\cala_R$ is a function $\tau$ on the positive cone\footnote{A positive operator $a \in \calb(\calh)$ obeys $\braket{x|a|x} \ge 0 \quad \forall \ket{x} \in \calh$. The space of positive operators is closed under linear combinations with non-negative coefficients and thus forms a cone. We write $a \ge 0$ to denote a positive operator. The trace can be extended from the positive cone to the full algebra by linearity.} $\cala_{R,+}$ valued in $[0,\infty]$ with the following properties:
\begin{enumerate}
\item $\tau(a + b) = \tau(a) + \tau(b), \quad a,b \in \cala_{R,+}$,
\item $\tau(\lambda a) = \lambda \tau(a), \quad \lambda \ge 0, \quad a \in \cala_{R,+}$,
\item $\tau(a^\dagger a) = \tau(a a^\dagger), \quad a \in \cala_{R}$,
\item $a \neq 0 \implies \tau(a) > 0, \quad a \in \cala_{R,+}, \quad$ (faithful)
\item $\forall a \in \cala_{R,+}, \, \exists b \in \cala_{R,+} \, \quad  a - b \ge 0, \quad \tau(b) < \infty, \quad b \neq 0, \quad$ (semifinite)
\item For every bounded increasing net $\{a_\alpha\}$ in $\cala_{R,+}$, $\tau(\sup a_\alpha) = \sup \tau(a_\alpha), \quad $ (normal),
\item $\tau$ respects the OPE identity \eqref{eq:4.52}, which can be verified given $\tau(a)$ for positive $a \in \cala_{R,0}$. To be precise, if we write the identity as $a = \sum_{n \ge 0} b_n$ for $a, b_n \in \cala_{R,0}$, then $\tau(a) = \sum_{n \ge 0} \tau(b_n)$.
\end{enumerate}
Property 7 is redundant, and has been included for our convenience. To argue for a unique trace on $\cala_R$, one invokes the following,\footnote{We have adapted this theorem from Corollary 2.32 of \cite{Takesaki1979} with the understanding that a type II$_\infty$ factor is semifinite (see Corollary 1.20 and Definition 1.21).}
\begin{theorem*}
On a type II$_\infty$ factor, any semifinite normal traces are proportional.
\end{theorem*}
It was shown in \cite{Penington:2023dql} that \eqref{eq:trace} has all of these properties. In the remainder of this section, we will present an alternate trace that is defined on the dense subalgebra $\cala_{R,0}$ and has properties 1, 2, 3, 4, 5, and 7.\footnote{For the purposes of applying these properties to a trace on $\cala_{R,0}$, we define the positive cone to be the space of operators that take the form $a^\dagger a$ for $a \in \cala_{R,0}$.} This trace is finite.\footnote{Had we included the identity in $\cala_{R,0}$, this would not be the case.} Due to the theorem, this densely-defined trace cannot be extended to a semifinite normal trace on all of $\cala_{R}$. However, the existence of such a trace is nontrivial due to property 7. Studying alternative densely-defined traces is interesting because corrections to the theory\footnote{JT gravity with matter can arise from a scaling or IR limit of a more general theory. Corrections to the bulk dynamics may be generated by backing away from the limit.} are expected to modify the algebra and the trace, and the modified trace could reasonably be expected to define a trace on $\cala_{R,0}$ that still respects property 7.\footnote{As we will explain shortly, the OPE identities may be verified in the asymptotic AdS region, and thus will be respected if the corrected theory behaves like JT gravity with matter at long distances.} Thus, by studying the space of densely defined traces that respect the OPE identities, one can constrain certain corrections to the bulk dynamics.

Given that \eqref{eq:trace} may be thought of as arising from the disk Euclidean path integral with a smooth interior geometry, a way to generate alternate traces is to introduce a defect in the interior of disk. For example, one could consider a conical defect or a geodesic hole. These were studied in \cite{Mertens:2019tcm}. Our alternate trace of interest corresponds to a cusp, which arises as a limiting case of both types of defects. To be precise, we let $T_b(a), \, a \in \cala_{R,0}, \, b > 0$ be the annulus (or trumpet) Euclidean path integral, where one boundary has the usual AdS boundary conditions and matter insertions as dictated by $a$, while the other boundary is a geodesic of length $b$. The matter state on this geodesic is prepared using a second trumpet path integral for the matter theory alone. In other words, the entire configuration is a double-trumpet at fixed $b$ where the boundary particle propagates along only one boundary. Our alternate trace is then defined as
\begin{equation}
\text{tr}_\lambda (a) := \text{tr}(a) + \lambda \lim_{b \rightarrow 0} \frac{T_b(a)}{\text{Tr}_{\calh_0} e^{-b H}}, \quad a \in \cala_{R,0}, \quad \lambda \ge 0,
\label{eq:alttrace}
\end{equation}
where $\lambda$ is arbitrary (hence this defines a 1-parameter family of traces, and $\text{tr}$ is recovered at $\lambda = 0$). We view the cusp path integral as a correction to the disk. We do not have a rigorous argument that $\text{tr}_\lambda$ for $\lambda > 0$ is finite on all of $\cala_{R,0}$, but it certainly is for the case that the matter theory is a free field. In this case, the $\text{Tr}_{\calh_0} e^{-b H}$ factor in the denominator cancels the determinant of the matter path integral on the double-trumpet, so the path integral may be evaluated using Wick contractions. Since our goal is to demonstrate that a weaker notion of the trace is no longer unique on $\cala_{R,0}$, the reader may choose to restrict their attention to a free matter field in this section only.

\begin{figure}
	\centering
	\includegraphics[width=0.3\linewidth]{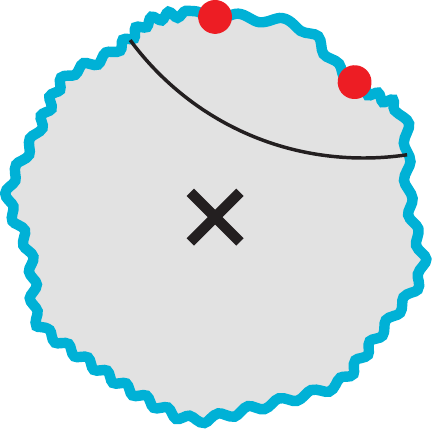}
	\caption{We depict the Euclidean disk two-point function in the presence of a generic localized bulk defect, denoted by the cross. The black line is a geodesic. The Euclidean path integral in the region above the geodesic prepares a state on the geodesic. This region is isomorphic to the half-disk, so the state belongs to $\calh$. An equivalent way to prepare this state is to replace the two operators by a sum over operators using the OPE identity \eqref{eq:4.52}. Hence, the disk path integral with the defect respects the OPE identities.}
	\label{fig:diskdefect}
\end{figure}

Given that $\text{tr}_\lambda$ is finite on $\cala_{R,0}$, we now argue that $\text{tr}_\lambda$ respects the OPE identities \eqref{eq:4.52}. See Figure \ref{fig:diskdefect}. The defect can be separated from the operator insertions using a geodesic, and the OPE identity may be verified by considering the Euclidean path integral in the region bounded by the geodesic, which prepares a state. Hence, the OPE identities may be verified locally on the AdS boundary.

From its representation as a Euclidean path integral with a defect, the alternate trace $\text{tr}_\lambda$ is clearly cyclic. The last property we need to prove is that this trace is positive and faithful. This is nontrivial because for $b > 0$, the linear functional $T_b$ is not positive. For example, from \eqref{eq:trumpet}, we have that
\begin{equation}
	T_b(e^{-\beta H_R}) = \frac{\text{Tr}_{\calh_0} e^{- b H}}{\pi} \int_0^\infty ds \, \cos(bs) e^{- \frac{s^2}{4 \phi_b} \beta}.
\end{equation}
The fact that $\cos(bs)$ takes on both positive and negative values for $s \in (0,\infty)$ indicates that there are positive functions of $H_R$ (which define positive operators) that are mapped to negative numbers by $T_b$. However, if we divide by $\text{Tr}_{\calh_0} e^{- b H}$ and then take $b \rightarrow 0$, we avoid this issue, as we now show.

We define the linear functional $\cali$ as follows:
\begin{equation}
	\cali(\cdot) :=	\int_0^\infty \frac{db}{2 \sinh \frac{b}{2}} T_b(\cdot).	
	\label{eq:calIdef}
\end{equation}
Earlier, we claimed that $T_b(\cdot) / \text{Tr}_{\calh_0} e^{-b H}$ is well-defined as $b \rightarrow 0$. Hence, \eqref{eq:calIdef} diverges due to the $b \rightarrow 0$ behavior of the integrand. We show in appendix \ref{sec:Iapp} that this divergence is always $+ \infty$ when a positive operator in $\cala_{R,0}$ is inserted into $\cali(\cdot)$. Hence, the second term in \eqref{eq:alttrace}, which is well-defined, must be nonnegative for positive $a$. Furthermore, even if the second term is zero, the first term will always be nonzero unless $a = 0$, because $\text{tr}$ is faithful. Hence, $\text{tr}_\lambda$ is faithful and positive for all $\lambda \ge 0$.

To summarize, we have defined a one-parameter family of traces on $\cala_{R,0}$ that respects the OPE identities and is cyclic, positive, and faithful. Hence, to arrive at the conclusion that the trace in \eqref{eq:trace} is unique, one must complete $\cala_{R,0}$ in the weak operator topology to obtain a vN algebra. Previous works that studied entanglement entropy in JT gravity with matter \cite{Lin:2021tlr,Goel:2018ubv}  have limited their attention to density matrices in $\cala_{R,0}$. Density matrices outside of $\cala_{R,0}$ but within $\cala_{R}$ are just as physical as those within $\cala_{R,0}$, so the Lorentzian theory only has one physically sensible notion of entropy, up to the $S_0$ ambiguity.

Note that the alternative trace $\text{tr}_\lambda$ may be used as an input to the GNS construction to obtain a new Hilbert space and new boundary algebras that differ from $\cala_R$ and $\cala_L$. In particular, $\text{tr}_\lambda$ defines a positive inner product on $\cala_{R,0}$, which may be completed to obtain a Hilbert space. It would be interesting to have an explicit Lorentzian description of this one-parameter family of theories. This is beyond the scope of the present work. In the next section, we discuss the representation theory of the boundary algebras, which is useful for addressing the factorization problem.

\section{The factorization problem}

\label{sec:facproblem}

In this section, we comment on the factorization problem in pure JT gravity and in JT gravity with matter. In particular, in \cite{Jafferis:2019wkd}, we factorized the Hilbert space of pure JT gravity and found a way to separately define the area and bulk entropy terms in the QES formula. It has been an open question as to whether one could similarly factorize the Hilbert space of JT gravity with matter, perhaps using edge modes, in a way that leads to a parallel understanding of the QES formula away from the semiclassical regime. We will argue that the algebraic structure of the boundary algebras prohibits such a factorization. Before doing so, we review some facts about the representations of type II$_1$ factors.

\subsection{Representations of the boundary algebras}

The goal of this section is to identify the possible representations of the boundary algebras. We will restrict our attention to $\cala_L$, but a similar discussion holds for $\cala_R$. Actually, let us consider the subalgebra that acts on the subspace of $\calh$ with left and right energies below some UV cutoff $\Lambda$. That is, define the projection $p_\Lambda \in \calb(\calh)$ by
 \begin{equation}
 p_\Lambda = \theta(\Lambda - H_L) \theta(\Lambda - H_R),
 \end{equation}
 where $\theta$ denotes the Heaviside step function. We then define $\cala^\Lambda_L := p_\Lambda \cala_L p_\Lambda$, which is a type II$_1$ vN factor on $\calh^\Lambda := p_\Lambda \calh$. Its commutant on $\calh^\Lambda$ is $\cala_R^\Lambda := p_\Lambda \cala_R p_\Lambda$. The identity operator on $\calh^\Lambda$ is $p_\Lambda$. Our discussion will be simpler with type II$_1$ algebras because they have a finite trace, which we still refer to as $\text{tr}$. We expect that our final conclusions will remain unchanged if we remove the UV cutoff.
 
In the literature, representations of vN algebras are also known as {\it modules}. We refer to \cite{popa} for the following definitions,
\begin{definition*}
A (left) module of a vN algebra $M$ is a Hilbert space $\calh$ with a normal unital homomorphism $\pi : M \rightarrow \calb(\calh)$.
\end{definition*}
\begin{definition*}
Let $M$ be a vN algebra that acts on the Hilbert space $\calh$, and let $N$ be a vN algebra on $\calk$. Let $M_+$ and $N_+$ denote their positive cones. A linear map $\Phi$ from $M$ into $N$ is {\it positive} if $\Phi(M_+) \subset N_+$. A positive linear map is {\it normal} if for every bounded increasing net $a_\alpha \in M_+$, we have $\Phi(\sup_\alpha a_\alpha) = \sup_\alpha \Phi(a_\alpha)$.
\end{definition*}

An example of a representation on $\cala_L^\Lambda$ is the {\it standard representation}. We let each element of $\cala_L^\Lambda$ denote a ket, and the inner product is defined by
\begin{equation}
\braket{ a | b } := \text{tr } a^\dagger b, \quad a,b, \in \cala_L^\Lambda.
\label{eq:7.2}
\end{equation}
This inner product is positive-definite because $\text{tr}$ is faithful. The Hilbert space, which we refer to as $L^2(\cala_L^\Lambda)$, is the completion of the space of kets with respect to the above inner product. We represent $\cala_L^\Lambda$ as follows:
\begin{equation}
\pi(a) \ket{b} := \ket{ab}.
\end{equation}
We say that $\cala^\Lambda_L$ acts on the Hilbert space by left-multiplication. We now argue that $L^2(\cala_L^\Lambda) = \calh^\Lambda$.\footnote{This is explained more thoroughly in section 3.3 of \cite{Penington:2023dql}.} Earlier, we pointed out that $\text{tr } a$ may be computed from the disk Euclidean path integral, where the boundary conditions are set by the operator $a$. We also showed that by dividing the boundary of the disk into an upper and lower half, we can rewrite the disk path integral as an overlap of a bra and a ket. The bra and ket may be separately computed using the Euclidean path integral on the half-disk, which prepares a state in the Hilbert space of JT gravity with matter. Imposing a UV cutoff $\Lambda$ does not affect this result. The bra and ket correspond to the left side of \eqref{eq:7.2}.  Hence, the standard representation of $\cala_L^\Lambda$ is the Hilbert space of JT gravity with matter quantized on a spatial interval with AdS boundary conditions, projected onto the sector obeying $H_L < \Lambda$ and $H_R < \Lambda$.

To see the relationship between $\cala_L^\Lambda$ and $\cala_R^\Lambda$ in the standard representation, let $J$ be the anti-unitary operator defined by
\begin{equation}
J \ket{a} := \ket{a^\dagger}.
\end{equation}
Next, note that
\begin{equation}
J a^\dagger J \ket{b} = \ket{ b a }.
\end{equation}
Because $\ket{a}$ corresponds to the Euclidean path integral on the half-disk with boundary conditions on the AdS boundary set by $a$, any operator in $\cala_R^\Lambda$ can be expressed as a right-multiplication in the standard representation. Hence, $\cala_R^\Lambda = J \cala_L^\Lambda J$.

The separable representations of a type II$_1$ factor are completely classified. From \cite{popa,jones,clare}, we learn that every separable representation of $\cala_L^\Lambda$ is equivalent to 
\begin{equation}
p(L^2(\cala_L^\Lambda) \otimes \ell^2(\mathbb{N})),
\end{equation}
where $\cala_L^\Lambda$ acts on $L^2(\cala_L^\Lambda)$ by left-multiplication, and $p$ is a projection in the commutant of $\cala_L^\Lambda$, which is $\cala_R^\Lambda \otimes \calb(\ell^2(\mathbb{N}))$, a type II$_\infty$ factor. This means that every representation of $\cala_L^\Lambda$ may be physically interpreted as the two-sided black hole described by $L^2(\cala_L^\Lambda)$ entangled with a bath.

\subsection{Comments on factorization}

\label{sec:factorization}

We now return to the question of whether the Hilbert space $\calh^\Lambda$ may be usefully factorized to obtain an alternate description of the QES formula \eqref{eq:heeformula} or even a realization of ER=EPR. The factorization map should be an isometric map from $\calh^\Lambda$ into
\begin{equation}
\label{eq:facmap}
\bigoplus_\alpha \calh_{\alpha,L} \otimes \calh_{\alpha,R},
\end{equation}
where $\alpha$ labels a superselection (or edge mode) sector, and $\calh_{\alpha,L}$ and $\calh_{\alpha,R}$ are representations of $\cala_L^\Lambda$ and $\cala_R^\Lambda$.\footnote{Instead of a direct sum, we also could have considered a direct integral, but the difference is not important for us.} Equation \eqref{eq:facmap} is inspired by successful attempts \cite{Donnelly:2014gva} to factorize the Hilbert space of two-dimensional Yang-Mills theory with compact gauge group $G$, where the Hilbert space is $L^2(G)$, and each $\alpha$ denotes a representation of $G$. The vN algebra associated with the spatial region to the left of the entangling surface is the group algebra that acts on $L^2(G)$ by left-multiplication \cite{Donnelly:2016auv}. This algebra is a direct sum of type I factors \cite{lurie}. In fact, any type I algebra acting on a Hilbert space may be represented using \eqref{eq:facmap}, where the algebra acts on the left tensor factors only \cite{Harlow:2016vwg}. Hence, a type I algebra induces a natural factorization map. The standard representation $L^2(\cala_L^\Lambda)$ appears morally similar to $L^2(G)$, because $\cala_L^\Lambda$ also acts on $L^2(\cala_L^\Lambda)$ by left-multiplication. However, the problem of factorizing $L^2(\cala_L^\Lambda)$ is different because $\cala_L^\Lambda$ is a type II$_1$ factor. As noted above, any representation of $\cala_L^\Lambda$ corresponds to a two-sided black hole, so $L^2(\cala_L^\Lambda)$ cannot be isometrically mapped into \eqref{eq:facmap} if $\calh_{\alpha,L}$ and $\calh_{\alpha,R}$ are to be interpreted as complementary entanglement wedges in the same two-sided spacetime. If we require that $\calh_{\alpha,L}$ and $\calh_{\alpha,R}$ furnish representations of $\cala_L^\Lambda$ and $\cala_R^\Lambda$, then \eqref{eq:facmap} can only be interpreted as a direct sum over pairs of entangled two-sided universes. Put differently, in \cite{Jafferis:2019wkd} we emphasized the importance of using a local boundary condition on a brick wall (or stretched horizon) to factorize the Hilbert space. If each tensor factor is a representation of a boundary algebra, then the only possible local boundary condition is a second AdS boundary.\footnote{On this boundary, one could insert operators in the commutant algebra. These would play the role of the defect operator.} As pointed out in section \ref{sec:pathintegralquant}, the trace over a single tensor factor is then equivalent to the double-trumpet Euclidean path integral, which is universally divergent. The reason why we were able to successfully factorize the Hilbert space in \cite{Jafferis:2019wkd} is because the commutative boundary algebras in pure JT gravity, which are generated by the Hamiltonian only, allow for a sensible factorization map.\footnote{The one-sided trace in our pure JT gravity analysis \cite{Jafferis:2019wkd} also produced a universal infinite multiplicative constant, owing to the continuous spectrum of the Hamiltonian. We interpreted this constant as the infinite volume of $\widetilde{SL}(2,\mathbb{R})$ and formally absorbed it into $S_0$ to obtain a finite result. We do not know how to formally absorb the universal divergence described in the main text into $S_0$.
}

\section{Discussion}

\label{sec:discussion}

In this paper, we studied the boundary algebras of JT gravity minimally coupled to an arbitrary matter QFT. This theory may be solved exactly for arbitrary values of the coupling, and the semiclassical limit is the weak-coupling limit. Our motivation is to better understand how semiclassical quantum gravity emerges from a more general theory. We defined the right boundary algebra $\cala_R$ to be the vN algebra generated by the right Hamiltonian and the matter operators on the right boundary. The left boundary algebra $\cala_L$ is defined analogously. We found that these two algebras are commutants of each other and that they are both factors. These facts together imply that the union of $\cala_R$ and $\cala_L$ generates the entire Hilbert space. One can thus draw an analogy between these two algebras and two copies of a holographic CFT at finite $N$. The difference is that in the holographic CFTs, the boundary algebras are type I, whereas the boundary algebras in JT gravity with matter are type II$_\infty$. Just as one can take the large $N$ limit of the holographic CFTs to obtain emergent vN algebras that describe subregions of a semiclassical spacetime \cite{Leutheusser:2021frk, Leutheusser:2021qhd, Leutheusser:2022bgi}, one can take the semiclassical limit of our boundary algebras. In particular, the semiclassical limits of $\cala_R$ and $\cala_L$ should correspond to the right and left entanglement wedges. Any operator in a semiclassical entanglement wedge should be expressible as a semiclassical limit of operators in the corresponding boundary algebra. This in principle provides an explicit means of understanding how bulk operators in the entanglement wedge may be reconstructed from the boundary CFT. For example, we expect that our setup can be used to define an asymptotically isometric code \cite{Faulkner:2022ada}. It would be interesting to investigate questions related to the complexity of reconstruction and the role of nonminimal QESs in this context \cite{Brown:2019rox}. In particular, \cite{Leutheusser:2022bgi} pointed out that the entanglement wedge can be defined algebraically by including modular-flowed single-trace operators in the emergent vN algebras. The alternate definition of the entanglement wedge, as the domain of dependence of a spatial region bounded by the minimal QES and the boundary subregion \cite{Engelhardt:2014gca}, should follow as a consequence. An important question is to find the analogous algebraic definition of a nonminimal QES. This should shed light on why bulge QESs contributed negatively to the Renyi 2-entropy in \cite{Chandrasekaran:2022eqq}.

There are various technical generalizations of our work that would be interesting to explore. For instance, one could include supersymmetry using the formalism developed in \cite{Lin:2022rzw,Lin:2022zxd}. One could also consider the boundary algebra of a single-sided black hole where spacetime ends on a single end-of-the-world (EoW) brane. The EoW brane theory without matter was studied in \cite{Gao:2021uro, Penington:2019kki}. The Hamiltonian for the theory with matter is given in \eqref{eq:eowop}, with $\mu$ replaced by $\mu + H$, where $\mu$ is the EoW brane tension and $H$ is the Hamiltonian of the matter theory quantized on half of global AdS$_2$. Because we interpret the boundary algebra as an entanglement wedge, we conjecture that the boundary algebra in this example is type I (in other words, the minimal QES is empty in the absence of a bath \cite{Penington:2019kki}).

Another generalization to consider is the $q$-deformation, which was studied in \cite{Berkooz:2018qkz,Lin:2022rbf,Goel:2023svz,Berkooz:2022mfk,Berkooz:2018jqr}. In particular, our diagrammatic rules should admit a $q$-deformation. In the case that the matter theory is a free scalar field, these $q$-deformed rules are known and are discussed in \cite{Berkooz:2018jqr,Jafferis:2022wez}. The single-particle sector in \cite{Lin:2022rbf} is analogous to a primary sector in our work. With an explicit formula for the $q$-deformed gravitationally dressed conformal blocks, one can consider the problem of bootstrapping all known solutions to the crossing equation. Our proofs in section \ref{sec:algebras} should generalize to $q$-deformed theories, as the diagrams that compute disk correlators look the same as their $q$-deformed counterparts.

It would also be interesting to show that the boundary algebras are hyperfinite. Any type II$_\infty$ factor may be expressed as a type II$_1$ factor times $\calb(\calh)$. We also  pointed out earlier that after imposing a UV cutoff, the boundary algebras become type II$_1$ factors. There is a unique hyperfinite type II$_1$ factor. Murray and von Neumann realized it as an infinite tensor product of entangled qubits \cite{mvn}. We expect that the boundary algebras are hyperfinite because JT gravity with matter can be realized as a large $N$ limit of a pair of entangled SYK models.

As we pointed out earlier, the algebraic structure of the boundary algebras prevents us from factorizing the Hilbert space in a way that parallels previous work in two-dimensional Yang-Mills or pure JT gravity \cite{Donnelly:2014gva,Blommaert:2018iqz,Lin:2018xkj}. This outcome is essentially a consequence of the locality of the matter QFT. To factorize the Hilbert space, one needs to modify the algebra to become type I. It would be interesting to study the minimal modifications that are necessary for a satisfactory factorization map to exist. These corrections should be nonlocal. A minimum requirement on the corrected algebra is that it obeys the OPE identities \eqref{eq:4.52}. Assuming our above conjecture for the EoW brane analysis, a candidate for such a corrected algebra is the one-sided boundary algebra in a spacetime that ends on an EoW brane. One must determine how the one-sided boundary algebra is produced from non-local corrections to the two-sided dynamics.

The boundary algebras $\cala_R$ and $\cala_L$ are not the only interesting vN algebras at strong-coupling. There are also type III$_1$ factors that describe Rindler wedges in the matter sector $\calh_0$  \eqref{eq:decomp}. These should give rise to the emergent type III$_1$ factors of \cite{Leutheusser:2021frk,Leutheusser:2021qhd} in the semiclassical limit. More generally, every vN algebra that can be defined on $\calh$ should be viewed as an emergent vN algebra \cite{Leutheusser:2022bgi} in the dual boundary theory (such as the SYK model in the appropriate scaling regime), because $\calh$ is the GNS Hilbert space.

In this work we considered algebras of operators dressed to either the left or right boundaries. It would be interesting to introduce an observer's worldline in the bulk (as in \cite{Chandrasekaran:2022cip, Witten:2023qsv}) and dress operators to that worldline to define a new algebra. This will be important for understanding de Sitter holography, where there are no spatial boundaries but there are emergent vN algebras.

\pagebreak

\paragraph{Acknowledgements}

I thank Scott Collier, Daniel Harlow, Daniel Jafferis, Adam Levine, Hong Liu, Ronak Soni, Jon Sorce, and Gabriel Wong for stimulating discussions. I also thank Daniel Harlow and Jon Sorce for comments on the draft. This work is supported by the U.S. Department of Energy, Office of Science, Office of High Energy Physics under grant Contract Number  DE-SC0012567 (High Energy Theory research) and the Department of Defense grant award KK2014.

	\appendix

\section{Double-trumpet correlator}

\label{sec:appendixa}

\begin{figure}
	\centering
	\includegraphics[width=0.86\linewidth]{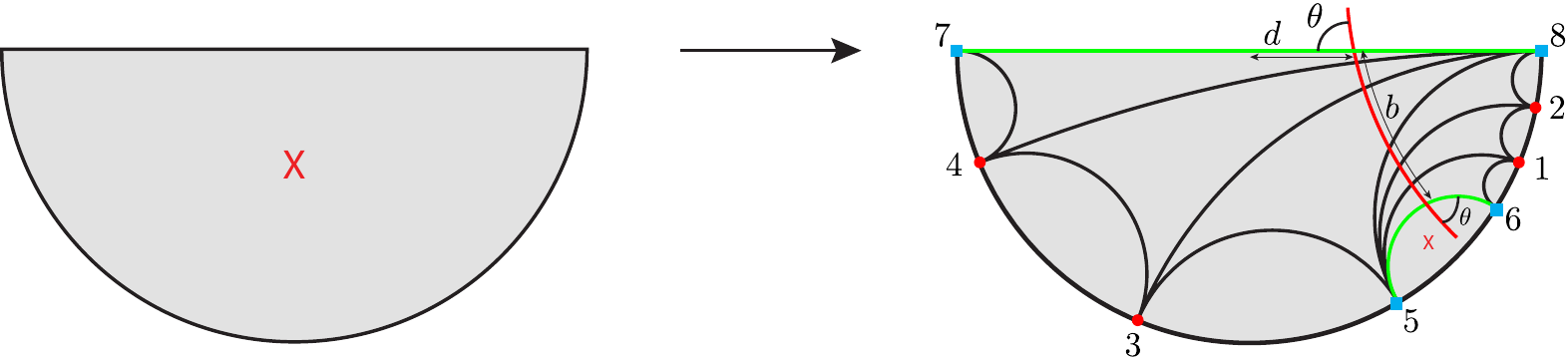}
	\caption{On the left, we depict a state in the matter theory prepared using the Euclidean path integral on the half-disk with an operator inserted somewhere. On the right, we show what becomes of this state after the action of the operator that appears inside the $\text{Tr}_{\calh_0}$ in equation \eqref{eq:bigintegral}. In particular, the entire half-disk on the left is mapped into the region bounded by the green geodesic connecting points 5 and 6 and the AdS boundary. Going forward, we will ignore this region and identify the two green geodesics, creating a double-trumpet geometry. All of the lines drawn are geodesics. The renormalized length between points $i$ and $j$ is $\ell_{ij}$. We will define $\ell := \ell_{78} = \ell_{56}$. The length of the red geodesic between the two green geodesics is $b$. Every double-trumpet has a unique minimal-length closed geodesic, which in this case is the red geodesic. We extended the red geodesic out of the fundamental domain so that we could easily label the angle $\theta$ that this geodesic makes with the green geodesics. The distance between the upper intersection point of the red and green geodesics and the polar coordinate origin of the half-disk is $d$. Given that the $7$-$8$ geodesic is centered about the origin, all of the points are fixed once the black and green geodesic lengths are specified. To be clear, each number refers to a specific point in the asymptotic infinity region of the disk that could be specified with a $(\phi,\psi)$ coordinate. Since these points are infinitely far away from the origin, we can only properly depict their $\phi$ coordinate in the figures.}
	\label{fig:busyfig}
\end{figure}

\begin{figure}
	\centering
	\includegraphics[width=0.7\linewidth]{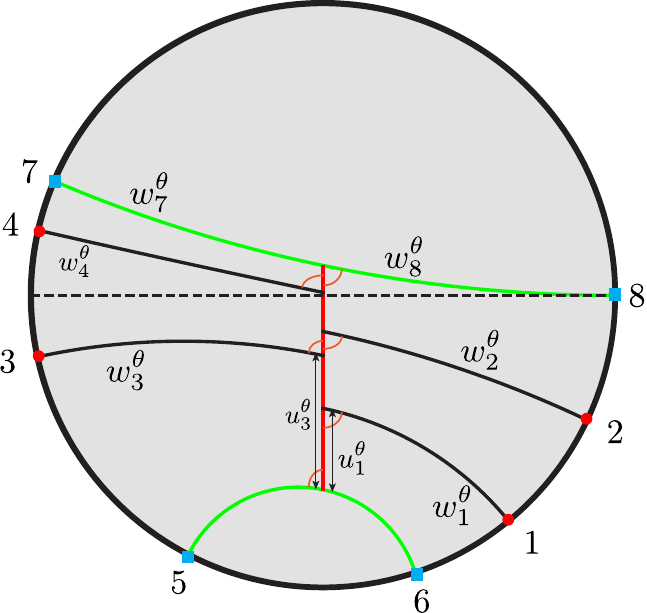}
	\caption{The region between the green geodesics is the same region between the green geodesics in Figure \ref{fig:busyfig}, but here we have chosen a reference frame where the red geodesic is oriented vertically and centered horizontally and point 8 has the angular polar coordinate $\phi = \pi$ (recall our conventions in Figure \ref{fig:coordconventions}). The renormalized length $w_1^\theta$ represents the distance along the black geodesic that connects point 1 with the red line at an angle of $\theta$. The $w^\theta_i$ coordinates, for $i \in \{2,3,4,7,8\}$, are defined similarly. All of the orange arcs denote an angle of $\theta$. The length of the red geodesic is $b$. The coordinate $u_1^\theta$ measures the distance between the bottom of the red geodesic and the intersection point of the aforementioned black geodesic and the red geodesic. Likewise, $u_3^\theta$ is defined analogously for point 3. We also wish to define $u_2^\theta$ and $u_4^\theta$, which are also measured from the bottom of the red geodesic, but we declined to draw them in the figure. The purpose of this figure is to define a coordinate system relative to the red geodesic, where one coordinate is measured along the red geodesic and the other coordinate is measured along a geodesic that makes an angle of $\theta$ with the red geodesic. We include the $\theta$ superscript because later we will consider a $\theta$-dependent change of coordinates.}
	\label{fig:lessbusyfig}
\end{figure}

Our goal is to compute
\begin{equation}
	\text{Tr } e^{-\dt \ell} e^{-\beta_6 H_L} \calo_L e^{-\beta_5 H_L} \calo_L e^{-\beta_4 H_L} e^{-\beta_3 H_R} \calo_R e^{-\beta_2 H_R} \calo_R e^{-\beta_1 H_R}.
\end{equation}
Using \eqref{eq:consisehr}, \eqref{eq:concisehl}, and \eqref{eq:orol}, we obtain
\begin{equation}
\begin{split}
&\int_{-\infty}^\infty 
d\ell \, d\ell_{15} \, d\ell_{25} \, d\ell_{58} \, d\ell_{16} \, d\ell_{12} \, d\ell_{28} \, d\ell_{35} \, d\ell_{38} \, d\ell_{34} \, d\ell_{48} \, d\ell_{47} \,
e^{-\dt \ell}
\\
&I(\ell,\ell_{15},\ell_{16})
I(\ell_{15},\ell_{25},\ell_{12})
I(\ell_{58},\ell_{25},\ell_{28})
I(\ell_{58},\ell_{35},\ell_{38})
I(\ell_{38},\ell_{34},\ell_{48})
I(\ell_{48},\ell,\ell_{47})
\\
&\Psi_{\beta_1}(\ell_{16}) \Psi_{\beta_2}(\ell_{12}) \Psi_{\beta_3}(\ell_{28})
\Psi_{\beta_4}(\ell_{35})
\Psi_{\beta_5}(\ell_{34})
\Psi_{\beta_6}(\ell_{47})
\\
&\text{Tr}_{\calh_0} \left[
e^{- e^{\frac{\ell_{47} - \ell_{48}}{2}} (H-B)} e^{ i P \frac{\ell_{48} - \ell}{2}}
\bdyO_L(\ell_{48}/2)
e^{- e^{\frac{\ell_{34} - \ell_{38}}{2}} (H-B)} e^{ i P \frac{\ell_{38} - \ell_{48}}{2}}
\bdyO_L(\ell_{38}/2)\right.
\\
&\left.e^{- e^{\frac{\ell_{35} - \ell_{58}}{2}} (H-B)} e^{ i P \frac{\ell_{58} - \ell_{38}}{2}}
e^{- e^{\frac{\ell_{28} - \ell_{25}}{2}} (H+B)} e^{- i P \frac{\ell_{25} - \ell_{85}}{2}}
\bdyO_R(\ell_{25}/2)
e^{- e^{\frac{\ell_{12} - \ell_{15}}{2}} (H+B)} e^{- i P \frac{\ell_{15} - \ell_{25}}{2}}
\right.
\\
&\left. \bdyO_R(\ell_{15}/2)
e^{- e^{\frac{\ell_{16} - \ell}{2}} (H+B)} e^{- i P \frac{\ell - \ell_{15}}{2}}
\right].	
\end{split}
\label{eq:bigintegral}
\end{equation}
We first address the operator that sits inside the trace in the matter sector, $\text{Tr}_{\calh_0}$. We represent the action of this operator in Figure \ref{fig:busyfig}, which was obtained by iterating the process illustrated in Figure \ref{fig:ehrfig} multiple times. The locations of all the labeled points in Figure \ref{fig:busyfig} are determined by the 12 renormalized geodesic lengths that are integrated over in \eqref{eq:bigintegral}. In Figure \ref{fig:lessbusyfig}, we introduce a more convenient coordinate system for the 12-dimensional domain of integration. The new coordinates are $\theta$, $b$, $w_i^\theta$ for $i \in \{1,2,3,4,7,8\}$, and $u_i^\theta$ for $i \in \{1,2,3,4\}$. The coordinate ranges are $(-\infty,\infty)$ with the following exceptions:
\begin{equation}
	b \in (0,\infty), \quad \theta \in (0,\pi), \quad 0 < u_1^\theta < u_2^\theta < b, \quad 0 < u_3^\theta < u_4^\theta < b. 
\label{eq:ranges}
\end{equation}
Note that the distance $d$ in Figure \ref{fig:busyfig} is equal to
\begin{equation}
	d = \frac{w_7^\theta - w_8^\theta}{2}.
\label{eq:d}
\end{equation}

Using this new coordinate system, we may obtain a new expression for the operator that appears within the $\text{Tr}_{\calh_0}$ in \eqref{eq:bigintegral}. First, we define the operator
\begin{equation}
	U_{d,\theta}(x) := \exp\left( x \left[B \sin \theta \, \sinh d  - i P \cos \theta - H \cosh d \, \sin \theta \right] \right),
\end{equation}
which corresponds to an isometry that moves the uppermost green geodesic in Figure \ref{fig:busyfig}, which has a $u^\theta$ coordinate of $b$, to a geodesic with a $u^\theta$ coordinate of $b - x$. Next, we define
\begin{equation}
	\bdyO_R^{(d,\theta)}(u,w) := U^{-1}_{d,\theta}(u) \bdyO_R(w) U_{d,\theta}(u), \quad \quad \bdyO_L^{(d,\theta)}(u,w) := U^{-1}_{d,\theta}(u) \bdyO_L(w) U_{d,\theta}(u).
\end{equation}
Finally, the operator that appears within the $\text{Tr}_{\calh_0}$ in \eqref{eq:bigintegral} is equal to
\begin{equation}
	\label{eq:simplerop}
	U_{d,\theta}(b) \calt_b \left\{  \bdyO_L^{(d,\theta)}(u_4^\theta,w_4^\theta - d)   \bdyO_L^{(d,\theta)}(u_3^\theta,w_3^\theta - d)      \bdyO_R^{(d,\theta)}(u_2^\theta,w_2^\theta + d)     \bdyO_R^{(d,\theta)}(u_1^\theta,w_1^\theta + d)      \right\},
\end{equation}
where we have introduced a special time-ordering signal $\calt_b$, which orders the operators based on their $u^\theta$ coordinates modulo shifts of $b$. That is, the $u^\theta$ coordinates should first be shifted by multiples of $b$ to fall in the range $(0,b)$, and then their order determines the order of the operators. Given the ranges \eqref{eq:ranges}, this procedure may seem pointless, but later we will consider other coordinate systems where the $u$ coordinate goes outside the $(0,b)$ range. To see why $\calt_b$ is useful, consider setting $u_4^\theta$ to $b$. Then the operator will be ordered first. Equivalently, one could have used cyclicity of the trace to move the operator to the front. Thus, the $\calt_b$ allows us to define the $u$ coordinates outside the range $(0,b)$ if we wish, with the understanding that the actual ordering is defined in the fundamental domain $(0,b)$.

Equation \eqref{eq:simplerop} represents a substantial simplification relative to the operator in \eqref{eq:bigintegral}. Because it appears in a trace, we can conjugate \eqref{eq:simplerop} to simplify it even further. If we conjugate by $e^{- i P d}$ then we can remove the $d$ dependence. The result is
\begin{equation}
	\label{eq:simpleropnod}
	U_{0,\theta}(b) \calt_b \left\{  \bdyO_L^{(0,\theta)}(u_4^\theta,w_4^\theta)   \bdyO_L^{(0,\theta)}(u_3^\theta,w_3^\theta)      \bdyO_R^{(0,\theta)}(u_2^\theta,w_2^\theta)     \bdyO_R^{(0,\theta)}(u_1^\theta,w_1^\theta)      \right\}.
\end{equation}

We now consider an infinitesimal, $\theta$-dependent change of coordinates on the $w^\theta$ and $u^\theta$ variables given by
\begin{equation}
	\label{eq:passivetransf}
	dw_i^\theta = - \cot \theta \, d\theta, \quad i \in \{1,2,3,4,7,8\}. \quad \quad du^\theta_i = 0, \quad i \in \{1,2\}. \quad \quad du^\theta_i = \frac{2 \, d\theta}{\sin \theta}, \quad i \in \{3,4\}. 
\end{equation}
Next, note that $U_{0,\theta}(b)$ obeys
\begin{align}
\frac{d U_{0,\theta}(b)}{d\theta} &= \left[i P \cot \theta + H - B,U_{0,\theta}(b)\right],
\end{align}
and using \eqref{eq:passivetransf}, we also have that
\begin{align}
	\frac{d}{d\theta} \bdyO_R^{(0,\theta)}(u_i^\theta,w_i^\theta) &= [i P \cot \theta + H - B, \bdyO_R^{(0,\theta)}(u_i^\theta,w_i^\theta)], \quad i \in \{1,2\}
	\\
		\frac{d}{d\theta} \bdyO_L^{(0,\theta)}(u_i^\theta,w_i^\theta) &= [i P \cot \theta + H - B, \bdyO_L^{(0,\theta)}(u_i^\theta,w_i^\theta)], \quad i \in \{3,4\}
\end{align}
It follows that changing $\theta$ in \eqref{eq:simpleropnod} by an infinitesimal amount $d\theta$, together with performing the infinitesimal transformation in \eqref{eq:passivetransf}, is equivalent to a conjugation, which does not affect the trace. We may thus replace \eqref{eq:simpleropnod} by
\begin{equation}
	\label{eq:simpleropnodaftertransf}
	e^{-b H} \calt_b \left\{  \bdyO_L^{(0,\frac{\pi}{2})}(u_4^{\frac{\pi}{2}},w_4^{\frac{\pi}{2}})   \bdyO_L^{(0,{\frac{\pi}{2}})}(u_3^{\frac{\pi}{2}},w_3^{\frac{\pi}{2}})      \bdyO_R^{(0,{\frac{\pi}{2}})}(u_2^{\frac{\pi}{2}},w_2^{\frac{\pi}{2}})     \bdyO_R^{(0,{\frac{\pi}{2}})}(u_1^{\frac{\pi}{2}},w_1^{\frac{\pi}{2}})      \right\},
\end{equation}
and we explicitly have that
\begin{equation}
	\label{eq:passivetransfintegrated}
	w_i^{\frac{\pi}{2}} = w_i^\theta + \log \sin \theta, \quad i \in \{1,2,3,4,7,8\}. \quad \quad  u^{\frac{\pi}{2}}_i= u^\theta_i, \quad i \in \{1,2\}. \quad \quad u^{\frac{\pi}{2}}_i = u_i^\theta + 2 \log \cot \frac{\theta}{2}, \quad i \in \{3,4\}. 
\end{equation}

\begin{figure}
	\centering
	\includegraphics[width=0.7\linewidth]{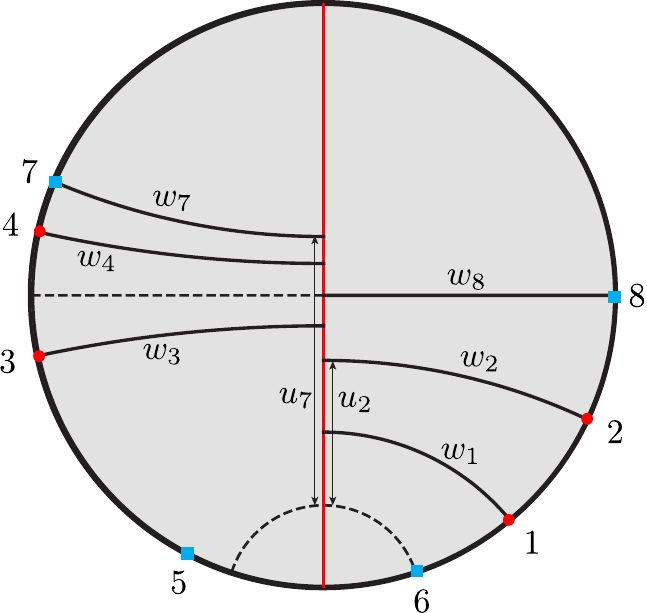}
	\caption{We display the final coordinate system that we are interested in. The locations of the numbered points are the same as in Figure \ref{fig:lessbusyfig}. The red geodesic has been extended indefinitely. Both the thick black lines and the dashed lines are geodesics that intersect the red geodesic at a right angle. A fundamental domain of the double-trumpet lies between the dashed geodesics (this fundamental domain differs from the one between the green geodesics in Figure \ref{fig:lessbusyfig}), and the length of the red geodesic between the two dashed geodesics is $b$. The lower dashed geodesic is related to the upper one by a vertical isometry (i.e. the corresponding operator in the matter theory would be $e^{-b H}$). The coordinates $w_i$, $i \in \{1,2,3,4,7,8\}$ measure the renormalized distance between a numbered point and the red geodesic. The coordinates $u_i$, $i \in \{1,2,3,4,7\}$ measure a distance along the red geodesic starting from where the red geodesic intersects the lower dashed line. We have only displayed $u_7$ and $u_2$ while suppressing the other $u_i$ coordinates. The six $w_i$ coordinates together with the five $u_i$ coordinates and $b$ comprise our final coordinate system for the domain of integration of \eqref{eq:bigintegral}. Note how points 6 and 8 intersect both the dashed geodesics here and the green geodesics in Figure \ref{fig:lessbusyfig} (the upper dashed geodesic is meant to extend all the way to the right but it is covered by a black geodesic).}
	\label{fig:lesslessbusyfig}
\end{figure}

To recapitulate our steps, we began with the 12 renormalized geodesic lengths in the integration measure of \eqref{eq:bigintegral}, and then changed these coordinates to $\theta$, $b$, $w_i^\theta$ for $i \in \{1,2,3,4,7,8\}$, and $u_i^\theta$ for $i \in \{1,2,3,4\}$. Then, we changed all of the $w_i^\theta$ and $u_i^\theta$ coordinates to $w_i^{\pi/2}$ and $u_i^{\pi/2}$ coordinates using \eqref{eq:passivetransfintegrated}. We now make one final change of coordinates: we exchange $\theta$ with $u_7^{\frac{\pi}{2}}$, defined by
\begin{equation}
	u_7^{\frac{\pi}{2}} := b + 2 \log \cot \frac{\theta}{2}.
\end{equation}
Finally, to ease the notation, \emph{we will drop the $\frac{\pi}{2}$ superscripts from now on.} The geometric interpretation of our final coordinate system can be understood from Figure \ref{fig:lesslessbusyfig}. The coordinates are $w_i$ for $i \in \{1,2,3,4,7,8\}$, $u_i$ for $i \in \{1,2,3,4,7\}$, and $b$. The coordinate ranges are
\begin{equation}
\label{eq:coordranges}
	w_i \in (-\infty,\infty), \quad b \in (0,\infty), \quad u_7 \in (-\infty,\infty), \quad u_7 - b < u_3 < u_4 < u_7, \quad 0 < u_1 < u_2 < b.
\end{equation}
The last two restrictions on the $u$ coordinates follow from the fact that the transformation in \eqref{eq:passivetransf} preserves the difference of any two such coordinates. The matter trace in \eqref{eq:bigintegral} in these coordinates is simple:
\begin{equation}
	\label{eq:simplemattertrace}
	\text{Tr}_{\calh_0} \left[ e^{-b H} \calt_b \left\{  \bdyO_L(u_4,w_4)   \bdyO_L(u_3,w_3 )      \bdyO_R(u_2,w_2)     \bdyO_R(u_1,w_1)      \right\}\right],
\end{equation}
where
\begin{equation}
	\bdyO_R(u,w) := e^{u H} \bdyO_R(w) e^{-u H}, \quad \quad \bdyO_L(u,w) := e^{u H} \bdyO_L(w) e^{-u H}.
\end{equation}
The coordinate transformation from the original 12 renormalized lengths is simple:
\begin{align}
	e^{-\ell_{ij}} &= 
\frac{
	e^{-(w_i + w_j)}
}{\left[\cosh\left(\frac{u_i - u_j}{2}\right)\right]^2}, \quad i \in \{3,4,5,7\}, \quad j \in \{1,2,6,8\}
\\
e^{-\ell_{ij}} &= 
\frac{
	e^{-(w_i + w_j)}
}{\left[\sinh\left(\frac{u_i - u_j}{2}\right)\right]^2}, \quad i,j \in \{3,4,5,7\}
\\
\\
e^{-\ell_{ij}} &= 
\frac{
	e^{-(w_i + w_j)}
}{\left[\sinh\left(\frac{u_i - u_j}{2}\right)\right]^2}, \quad i,j \in \{1,2,6,8\}
\end{align}
Above, $w_5$ should be replaced by $w_7$, $w_6$ should be replaced by $w_8$, $u_8$ should be replaced with $b$, $u_6$ should be replaced with $0$, and $u_5$ should be replaced with $u_7 - b$. Also, $\ell_{78} = \ell_{56} = \ell$.

After performing the change of integration variables, \eqref{eq:bigintegral} finally becomes
	\begin{align}
		\label{eq:ifirst}
	&\int  db \,   du_7 \, \left( \frac{e^{\frac{-w_7 - w_8}{2}}}{\cosh\left( \frac{u_7 - b}{2}\right)} \right)^{2 \dt} \, \text{Tr}_{\calh_0} \, \biggl[ e^{-b H} \calt_b \biggl\{
\\[20pt]	
\label{eq:isecond}
\begin{split}
&2 \sinh \frac{b}{2} \frac{du_1 dw_1 e^{w_1}}{2} \frac{du_2 dw_2 e^{w_2}}{2} \frac{dw_8 e^{w_8}}{2}
\\
&\calk^t_{\beta_3}(b,w_8;u_2,w_2)
\bdyO_R(u_2,w_2)
\calk^t_{\beta_2}(u_2,w_2;u_1,w_1)
\bdyO_R(u_1,w_1)
\calk^t_{\beta_1}(u_1,w_1;0,w_8)
\end{split}
\\[10pt]
\label{eq:ithird}
\begin{split}
&\,
	\\
	&2 \sinh \frac{b}{2} \frac{du_3 dw_3 e^{w_3}}{2} \frac{du_4 dw_4 e^{w_4}}{2} 	\frac{ dw_7 e^{w_7}}{2} 
\\
&\calk^t_{\beta_6}(u_7,w_7;u_4,w_4)
\bdyO_L(u_4,w_4)
\calk^t_{\beta_5}(u_4,w_4;u_3,w_3)
\bdyO_L(u_3,w_3)
\calk^t_{\beta_4}(u_3,w_3;u_7-b,w_7) \biggr\} \biggr]
\end{split}
	\end{align}
The way we wrote the integrand is not meant to imply any particular order of integration. The domain of integration is provided in \eqref{eq:coordranges}. Line \eqref{eq:isecond} is the amplitude for the boundary particle to propagate around a trumpet, starting at point 6 and returning to point 8, which is identified with point 6. There are also matter operators inserted. One may compare with \eqref{eq:trumpetcalc}, which is similar but does not have matter operators. As discussed in section \ref{sec:trumpet}, the $2 \sinh \frac{b}{2}$ factor ensures that the measure of the double-trumpet wiggles is correct. Line \eqref{eq:ithird} is similar, but for the other trumpet. Note that the start and end points of the boundary particle depends on $u_7$, which is integrated from $-\infty$ to $\infty$. Hence, we are integrating over a relative bulk Euclidean global time translation between the two boundaries. When $u_7$ is shifted by $b$, the start/end point of the boundary particle on the trumpet is the same, but the geodesic that connects this point with a fixed point on the other boundary is wound once around the double-trumpet. Hence, this integral computes the double-trumpet amplitude with matter insertions on the boundaries and also with a sum over all geodesics that connect the two boundaries at specific marked points. This sum is infinite due to the infinitely many ways a geodesic can wind around the double-trumpet, and the term in \eqref{eq:ifirst} involving $\dt$ regulates this sum. Note that the integral is still divergent because as $b$ approaches zero, the matter path integral diverges in unitary matter theories with local degrees of freedom.

\section{A complete basis of $\calh$}
\label{sec:spanning}
In this appendix, we show that states of the form \eqref{eq:HHmatter} span the entire Hilbert space $\calh$. We also find an orthonormal basis of $\calh$ (namely the energy basis). To do this, we start with \eqref{eq:HHmatter} but perform an inverse Laplace transform on $\beta_1$ and $\beta_2$. The result is
\begin{equation}
	\begin{split}
		&\ket{\Psi^{\bdyO}_{s_1,s_2}(\ell)} :=
		\int_{-\infty}^\infty d\psi_1 \int_0^\pi d\phi_1
		e^{\Delta \psi_1 } \ket{\bdyO(\phi_1) }
		\\
		& e^{-2(e^{\psi_0} + e^{\psi_1})\csc \phi_1} \rho(s_1) 2K_{2 i s_1}\left(2 e^{\frac{\psi_0 + \psi_1}{2}} \sec \frac{\phi_1}{2}\right)
		\rho(s_2) 2K_{2 i s_2}\left(2 e^{\frac{\psi_0 + \psi_1}{2}} \csc \frac{\phi_1}{2}\right) \csc \phi_1
		\biggr|_{\psi_0 =  \log 2 - \frac{\ell}{2}}.
	\end{split}
\end{equation}
These states diagonalize $H_L$ and $H_R$. Next, we use \eqref{eq:bessel} to go to a fixed length basis,
\begin{equation}
	\begin{split}
		&\ket{\Psi^{\bdyO}_{\ell_1,\ell_2}(\ell)} :=
		\int_{-\infty}^\infty d\psi_1 \int_0^\pi d\phi_1
		e^{\Delta \psi_1 } \ket{ \bdyO(\phi_1) }
		\\
		&e^{-2(e^{\psi_0} + e^{\psi_1})\csc \phi_1} \delta\left(\ell_2 + 2 \log \frac{1}{2} e^{\frac{\psi_0 + \psi_1}{2}} \sec \frac{\phi_1}{2}\right)
		\delta\left(\ell_1 + 2 \log \frac{1}{2} e^{\frac{\psi_0 + \psi_1}{2}} \csc \frac{\phi_1}{2}\right) \csc \phi_1
		\biggr|_{\psi_0 =  \log 2 + \frac{\ell}{2}}.
	\end{split}
\end{equation}
The delta functions become
\begin{equation}
	\delta\left(\ell_2 + 2 \log \frac{1}{2} e^{\frac{\psi_0 + \psi_1}{2}} \sec \frac{\phi_1}{2}\right)
	\delta\left(\ell_1 + 2 \log \frac{1}{2} e^{\frac{\psi_0 + \psi_1}{2}} \csc \frac{\phi_1}{2}\right) = \frac{\sin \phi_1}{2} \delta(\phi_1 - \phi_1^{\ell_1,\ell_2}) \delta(\psi_1 - \psi_1^{\ell_1,\ell_2})
\end{equation}
where
\begin{equation}
	\tan \frac{\phi_1^{\ell_1,\ell_2}}{2} := e^{\frac{\ell_1 - \ell_2}{2}}, \quad \psi_1^{\ell_1,\ell_2} := -\psi_0 - \log\left(\left[e^{\ell_1} + e^{\ell_2}\right]/4\right).
\end{equation}
We have that
\begin{equation}
	\begin{split}
		&\ket{\Psi^{\bdyO}_{\ell_1,\ell_2}(\ell)} =
		\frac{1}{2}
		\left(\frac{2 e^{-\ell/2}}{e^{\ell_1} + e^{\ell_2}}\right)^\Delta e^{-4( e^{\ell/2} + \frac{ e^{-\ell/2}}{e^{\ell_1} + e^{\ell_2}})\csc \phi_1^{\ell_1,\ell_2}} \, \ket{ \bdyO(\phi_1^{\ell_1,\ell_2}) }.
	\end{split}
\end{equation}
By taking arbitrary linear combinations of the above with $\ell_1 - \ell_2$ fixed, we can produce an arbitrary function of $\ell$ times $\ket{ \bdyO(\phi_1^{\ell_1,\ell_2}) }$ for some fixed angle $\phi_1^{\ell_1,\ell_2}$. Thus, by taking suitable linear combinations, we can produce an arbitrary function of $\ell$ times an arbitrary descendant of $\bdyO(\frac{\pi}{2})$. Hence, states of the form $\ket{\Psi^\bdyO_{\beta_1,\beta_2}(\ell)}$ (where $\bdyO$ ranges over all primary operators) span $\calh$. Furthermore, note that the states $\ket{\Psi^{\bdyO}_{s_1,s_2}(\ell)}$ and $\ket{\Psi^{\tilde{\bdyO}}_{\tilde{s}_1,\tilde{s}_2}(\ell)}$ are orthogonal for $s_1 \neq \tilde{s_1}$, $s_2 \neq \tilde{s_2}$, or $\bdyO \neq \tilde{\bdyO}$ (we are assuming that the two-point function of distinct primaries vanishes). Hence, the states $\ket{\Psi^{\bdyO}_{s_1,s_2}(\ell)}$ form an orthogonal basis of $\calh$. Note the special case where $\bdyO$ is the identity operator. For this special case, the left and right energies are constrained to be equal, and the corresponding subspace of $\calh$ is identical to the Hilbert space of pure JT gravity.

\section{Non-negativity of $H + B$}
\label{sec:positive}
In this section we show that $H+B$ is a non-negative operator on $\calh_0$, the matter Hilbert space. Because the matter Hilbert space is a direct sum of discrete series representations, we simply need to check that $H+B$ is a non-negative operator in each representation. Each representation has a lowest-weight state $\ket{\Delta;0}$ that obeys
\begin{equation}
	L_1 \ket{\Delta;0} = 0, \quad L_0 \ket{\Delta;0} = \Delta \ket{\Delta;0}.
\end{equation}
We act on $\ket{\Delta;0}$ with $L_{-1}$ to obtain other states. In particular, we have that 
\begin{align} L_- \ket{\Delta ; n} &= \sqrt{(n + 2 \Delta)(n + 1)} \ket{\Delta ; n+ 1}, \\
 L_+ \ket{\Delta ; n } &= \sqrt{(2 \Delta + n - 1)n} \ket{\Delta ; n - 1},
 \end{align}
where $n \in \mathbb{Z}_{\ge 0}$ above. Also, unitarity implies $\Delta > 0$. Note that $H+B = L_0 + \frac{L_+ + L_-}{2}$. We define an eigenstate of $H+B$ with eigenvalue $\lambda$ as follows:
\begin{equation}
	\ket{\Delta;\lambda} := \sum_{n=0}^\infty\ket{\Delta;n} \frac{2^\Delta e^{-\lambda} \lambda^{\Delta - \frac{1}{2}} U(-n,2\Delta,2 \lambda)}{\sqrt{n! \Gamma(2 \Delta) (2 \Delta)_n}}, 
\end{equation}
where we have used the \href{https://reference.wolfram.com/language/ref/HypergeometricU.html}{HypergeometricU} function in Mathematica. One can explicitly check that $\ket{\Delta ; \lambda}$ is an eigenstate of $H+B$. Furthermore, note that
\begin{equation}
\int_0^\infty d\lambda	\braket{\Delta;n|\Delta;\lambda} \braket{\Delta;\lambda|\Delta;m} = \delta_{nm},
\end{equation}
so these eigenstates of $H+B$ form a complete basis. Because $\lambda$ only takes non-negative values above, it follows that $H+B$ is a non-negative operator. Similarly, $H-B$ is a non-negative operator.

\section{Positivity of alternate densely-defined trace}

\label{sec:Iapp}

Our goal is to show that the linear functional $\cali$, defined in \eqref{eq:calIdef}, is always $+\infty$ for positive operators in $\cala_{R,0}$. Explicitly, we have for example
\begin{equation}
	\label{eq:iexp}
	\begin{split}
		&\cali(e^{-\beta_3 H_R} \calo_R e^{-\beta_2 H_R} \calo_R e^{-\beta_1 H_R}) = \int_0^\infty\frac{db}{\text{Vol } U(1)}  \int \frac{du_1 dw_1 e^{w_1}}{2} \frac{du_2 dw_2 e^{w_2}}{2} \frac{du_3 dw_3 e^{w_3}}{2}
		\\
		&\text{Tr}_{\calh_0} \left[ e^{-b H} \calk^t_{\beta_3}(u_3 + b,w_3;u_2,w_2)
		\bdyO_R(u_2,w_2)
		\calk^t_{\beta_2}(u_2,w_2;u_1,w_1)
		\bdyO_R(u_1,w_1)
		\calk^t_{\beta_1}(u_1,w_1;u_3,w_3) \right].
	\end{split}
\end{equation}
Next, we rewrite \eqref{eq:iexp} by exchanging the variables $u_1$, $u_2$, $u_3$, and $b$ for the variables $u_{13} := u_1 - u_3$, $u_{21} := u_2 - u_1$, $u_{32} := u_3 + b - u_2$,
\begin{equation}
	\label{eq:iexp2}
	\begin{split}
		\cali(e^{-\beta_3 H_R} \calo_R e^{-\beta_2 H_R} \calo_R e^{-\beta_1 H_R}) =   \int_{-\infty}^\infty dw_1 &dw_2 dw_3 
		\\
		\text{Tr}_{\calh_0} \biggl[ &\left(\int_0^\infty \frac{du_{32}}{2}\,  e^{-u_{32} H} e^{\frac{w_3}{2}}\calk^t_{\beta_3}(u_{32},w_3;0,w_2) e^{\frac{w_2}{2}}\right)
		\bdyO_R(w_2) 
		\\
		&\left(\int_0^\infty \frac{du_{21}}{2} \, e^{-u_{21} H}
		e^{\frac{w_2}{2}}\calk^t_{\beta_2}(u_{21},w_2;0,w_1) e^{\frac{w_1}{2}} \right) 
		\bdyO_R(w_1) 
		\\
		&\left( \int_0^\infty \frac{du_{13}}{2} \, e^{-u_{13} H}
		e^{\frac{w_1}{2}}\calk^t_{\beta_1}(u_{13},w_1;0,w_3) e^{\frac{w_3}{2}} \right)  \biggr].
	\end{split}
\end{equation}
Next, we use 6.647.1 of \cite{grad} to calculate that
\begin{equation}
	\int_0^\infty \frac{du}{2} \, e^{-u H}
	e^{\frac{w_2}{2}}\calk^t_{\beta}(u,w_2;0,w_1) e^{\frac{w_1}{2}} = \int_0^\infty ds \, \rho(s) \, e^{-\frac{s^2}{4 \phi_b} \beta}  \, \calw_{-H,is}(w_2) \calw_{-H,is}(w_1)
\end{equation}
where
\begin{equation}
	\calw_{-\mu,is}(w) := \frac{e^{\frac{w_2}{2}}}{2} \left|\Gamma\left(\frac{1}{2} + i s + \mu\right)\right| W_{- \mu,is}(4 e^{-w}), \quad \mu > - \frac{1}{2}, \quad s \in \mathbb{R},
\end{equation}
and $W_{- \mu,is}(4 e^{-w})$ is the Whittaker W-function, which is real for the parameter ranges specified above. Note that
\begin{equation}
	\int_{-\infty}^\infty dw\, \calw_{-\mu,is_1}(w) \calw_{-\mu,is_2}(w) = \frac{\delta(s_1-s_2)}{\rho(s_1)}, \quad \int_{0}^\infty ds \, \rho(s) \, \calw_{-\mu,is}(w_1) \calw_{-\mu,is}(w_2) = \delta(w_1 - w_2).
	\label{eq:identities}
\end{equation}
In particular, note that $\calw_{-\mu,is}(w)$ is an eigenfunction of the following operator on $L^2(\mathbb{R})$,
\begin{equation}
	{\bf H_\mu} := 	\frac{1}{2 \phi_b}\left[ - \frac{1}{2} \partial_w^2 + 2 \mu e^{-w} + 2 e^{-2 w}\right],
	\label{eq:eowop}
\end{equation}
with eigenvalue $\frac{s^2}{4 \phi_b}$. In fact, \eqref{eq:eowop} is the Hamiltonian of pure JT gravity quantized on the spatial interval with one AdS boundary and one end-of-the-world brane boundary with tension $\mu$ \cite{Gao:2021uro}. This theory is described by the quantum mechanics of a particle scattering in the potential
\begin{equation}
	V_\mu(w) = 2 \mu e^{-w} + 2 e^{-2 w},
\end{equation}
and for $\mu > - \frac{1}{2}$ there are only scattering states and no bound states. Thus, for $\mu > - \frac{1}{2}$, the wavefunctions of the scattering states $\calw_{-\mu,is}(w)$ obey \eqref{eq:identities}. It follows that
\begin{equation}
	\cali(e^{-\beta_3 H_R} \calo_R e^{-\beta_2 H_R} \calo_R e^{-\beta_1 H_R}) = \text{Tr}_{\calh}\left[ e^{-\beta_3 {\bf H}_H} \calo_R e^{-\beta_2 {\bf H}_H} \calo_R e^{-\beta_1 {\bf H}_H}  \right].
\end{equation}
Because we can write $\cali$ as a trace on $\calh$, $\cali$ is a positive linear functional. In fact, it is always $+\infty$, which is what we wanted to show.

\section{Three-point function}

\label{sec:threepointapp}

In this appendix, we simplify \eqref{eq:4.49}, which we repeat here:
\begin{equation}
	\begin{split}
	& \sqrt{\Gamma^{\Delta_i}_{s_1,s_2} \Gamma^{\Delta_j}_{s_2,s_3} \Gamma^{\Delta_k}_{s_3,s_1} } \,  V_{ijk}(s_1,s_2,s_3)
	\\
		&= \int_{-\infty}^\infty d\ell_{ij} \, d\ell_{jk} \, d\ell_{ki} \, \, 
		2K_{2 i s_2}(4e^{-\ell_{ij}/2})
		2K_{2 i s_3}(4e^{-\ell_{jk}/2})
		2K_{2 i s_1}(4e^{-\ell_{ki}/2})
	    I(\ell_{ij},\ell_{jk},\ell_{ki})
	\\
	&\quad \times C_{ijk} \left(2 e^{-\ell_{ij}/2}\right)^{\Delta_i + \Delta_j - \Delta_k} \left(2 e^{-\ell_{ki}/2}\right)^{\Delta_i + \Delta_k - \Delta_j} \left(2 e^{-\ell_{jk}/2}\right)^{\Delta_j + \Delta_k - \Delta_i}.
	\end{split}
\label{eq:E1}
\end{equation}
For simplicity, we define
\begin{equation}
\Delta_a := \Delta_i + \Delta_j - \Delta_k, \quad \Delta_b := \Delta_j + \Delta_k - \Delta_i, \quad \Delta_c := \Delta_i + \Delta_k - \Delta_j, \quad .
\end{equation}
Given that $\Delta_i$, $\Delta_j$, and $\Delta_k$ are positive, it follows that at least two of $\Delta_a$, $\Delta_b$, and $\Delta_c$ are positive. We assume w.l.o.g. that $\Delta_a$ and $\Delta_c$ are positive. We will temporarily assume that $\Delta_b$ is also positive (so that we may use \eqref{eq:besselidentity} below), and then obtain an expression that may be analytically continued to negative values of $\Delta_b$.

We use the second line of \eqref{eq:Idef} as well as \eqref{eq:besselidentity} to obtain
\begin{equation}
	\begin{split}
	& \sqrt{\Gamma^{\Delta_i}_{s_1,s_2} \Gamma^{\Delta_j}_{s_2,s_3} \Gamma^{\Delta_k}_{s_3,s_1} } \,  V_{ijk}(s_1,s_2,s_3) =  C_{ijk}  \int_0^\infty ds \, \rho(s) \, \Gamma^{\frac{\Delta_a}{2}}_{s_2,s} \Gamma^{\frac{\Delta_b}{2}}_{s_3,s} \Gamma^{\frac{\Delta_c}{2}}_{s_1,s}.
	\end{split}
\label{eq:E2}
\end{equation}
Using (2.3) of \cite{neretin}, this becomes
\begin{equation}
	\begin{split}
		& \sqrt{\Gamma^{\Delta_i}_{s_1,s_2} \Gamma^{\Delta_j}_{s_2,s_3} \Gamma^{\Delta_k}_{s_3,s_1} } \,  V_{ijk}(s_1,s_2,s_3) 
		\\
		&= C_{ijk} \frac{\Gamma(\frac{\Delta_a+ \Delta_b}{2} + i s_2 \pm i s_3)  \Gamma(\frac{\Delta_a + \Delta_c}{2} - i s_2 \pm i s_1)  }{\Gamma(\Delta_a)}   
		\\
		&\times \int_{- i \infty + r}^{i \infty + r} \frac{dt}{2 \pi i}  
		\frac{\Gamma(\Delta_a + t) \Gamma(\frac{\Delta_a + \Delta_c}{2} + i s_2 \pm i s_1 + t) \Gamma(\frac{\Delta_b - \Delta_a}{2} \pm i s_3 - i s_2 - t) \Gamma(-t)}{\Gamma(\Delta_a + \Delta_c + t)\Gamma(\Delta_b - t) }.
	\end{split}
	\label{eq:E4}
\end{equation}
We may use a contour where $t$ has a fixed real part $r$ greater than $- \min(\Delta_a,\Delta_i)$ and less than $\min(0,\Delta_k - \Delta_i)$. This expression makes sense even when $\Delta_b$ is negative. For example, let us set $\Delta_i = \Delta_j + \Delta_k + n$, such that $\Delta_b = - n$. Let $n \in \mathbb{Z}_{\ge 0}$. Then \eqref{eq:E4} becomes
\begin{equation}
	\begin{split}
		& \sqrt{\Gamma^{\Delta_j + \Delta_k + n}_{s_1,s_2} \Gamma^{\Delta_j}_{s_2,s_3} \Gamma^{\Delta_k}_{s_3,s_1} } \,  V_{ijk}(s_1,s_2,s_3) 
		\\
		&= C_{ijk} \frac{\Gamma(\Delta_j + i s_2 \pm i s_3)  \Gamma(\Delta_j + \Delta_k + n - i s_2 \pm i s_1)  }{\Gamma(2 \Delta_j + n)}   
		\\
		&\times \int_{- i \infty + r}^{i \infty + r} \frac{dt}{2 \pi i}  
		\frac{\Gamma(2 \Delta_j + n + t) \Gamma(\Delta_j + \Delta_k + n + i s_2 \pm i s_1 + t) \Gamma(- \Delta_j - n \pm i s_3 - i s_2 - t) \Gamma(-t)}{\Gamma(2 \Delta_j + 2 \Delta_k + 2 n + t)\Gamma(-n - t) },
	\end{split}
	\label{eq:E5}
\end{equation}
which evaluates to
\begin{equation}
\begin{split}
&\sqrt{\Gamma^{\Delta_j + \Delta_k + n}_{s_1,s_2} \Gamma^{\Delta_j}_{s_2,s_3} \Gamma^{\Delta_k}_{s_3,s_1} } \,  V_{ijk}(s_1,s_2,s_3)
\\
&=C_{ijk} \sqrt{ \frac{ n! (n + 2 \Delta_j + 2 \Delta_k - 1) (2 \Delta_j + 2 \Delta_k + n)_n}{ (2 n + 2 \Delta_j + 2 \Delta_k - 1) (2 \Delta_j)_n (2 \Delta_k)_n } } \sqrt{\Gamma^{\Delta_j + \Delta_k + n}_{s_1,s_2} \Gamma^{\Delta_j}_{s_2,s_3} \Gamma^{\Delta_k}_{s_3,s_1} } P^{\Delta_k,\Delta_j}_n(s_3 ; s_1,s_2),
\end{split}
\end{equation}
where $P^{\Delta_k,\Delta_j}_n(s_3 ; s_1,s_2)$ was defined in (A.7) of \cite{Jafferis:2022wez}. We finally have that for $\Delta_i = \Delta_j + \Delta_k + n$, $n \in \mathbb{Z}_{\ge 0}$,
\begin{equation}
V_{ijk}(s_1,s_2,s_3) = C_{ijk} \sqrt{ \frac{ n! (n + 2 \Delta_j + 2 \Delta_k - 1) (2 \Delta_j + 2 \Delta_k + n)_n}{ (2 n + 2 \Delta_j + 2 \Delta_k - 1) (2 \Delta_j)_n (2 \Delta_k)_n } }  P^{\Delta_k,\Delta_j}_n(s_3 ; s_1,s_2).
\end{equation}
We can use this expression to compute a four-point conformal block of identical external primary operators of dimension $\Delta$, where the exchanged operator has dimension $2 \Delta + n$. We draw the rightmost diagram in Figure \ref{fig:fourptfig} and apply the diagrammatic rules described in the main text. We omit the $C_{ijk}$ factor because we only want to compute a conformal block. From (C.1) of \cite{Jafferis:2022wez}, we see that the semiclassical limit of the expression is indeed the conformal block.

	\bibliographystyle{JHEP}
	\nocite{}
	\bibliography{thebibliography}

\end{document}